\documentclass[intlimits,twoside,a4paper]{article}

\usepackage[cp1251]{inputenc}

\usepackage[eqsecnum]{cmpj3}

\usepackage{bm}

\issue{2019}{22}{3}{33702}
\doinumber{10.5488/CMP.22.33702}

\title[The monoclinic Ba$_2$P$_7$X (X $=$ Cl, Br, I) Zintl]%
{Investigation of structural
and elastic properties of monoclinic Ba$_2$P$_7$X (X $=$ Cl, Br, I) Zintl Salts compounds%
}
\author[M. Radjai \textsl{et al.}]{M. Radjai\refaddr{label1,label2}\thanks{Missoum RADJAI: E-mail: mradjai@yahoo.com},
        D. Maouche\refaddr{label1}, N. Guechi\refaddr{label3}, S. Cheddadi\refaddr{label4}, Z. Kechidi\refaddr{label5}}
\addresses{
\addr{label1} Laboratory for Developing New Materials and Their Characterizations, University Ferhat Abbas Setif 1, Algeria
\addr{label2} Laboratory of Physics of Experimental Techniques and Their Applications (LPTEAM), University of Medea, Algeria 
\addr{label3} Laboratory of Surfaces and Interfaces Studies of Solid Materials, University Ferhat Abbas Setif 1, Algeria
\addr{label4} Laboratory of Radiation Physics, Department of Physics, Faculty of Sciences, University Badji Mokhtar, Annaba, Algeria
\addr{label5} Laboratory of Electrical Engineering and Automatics LREA, University of Medea, Algeria
}

\date{Received	April 28, 2019, in final form June 21, 2019}

\begin{document}

\maketitle

\begin{abstract}
Structural and elastic properties of Ba$_{2}$P$_{7}$X (X=Cl, Br, I) ( Barium
Phosphide Halides) Zintl compounds have been investigated using the pseudo-potential plane-wave (PP-PW) method based on the density functional theory (DFT) within the generalized gradient approximation (GGA-PBESOL). The calculated lattice constants and internal parameters are in a good agreement with the experimental results reported in literature. In this paper, we present an investigation of the relative changes of the structural parameters and elastic constants as function of hydrostatic pressure. Isotropic elastic moduli and their related properties for single-crystal and polycrystalline phase, including the namely bulk modulus, shear modulus,
Young's modulus, Poisson's ratio, elastic anisotropy indexes, Pugh's indicator of brittle/ductile behavior, elastic wave velocities and Debye temperature have been estimated from $C_{ij}$ using Voigt, Russ and Hill approximations. Two different methods have been used to study the elastic anisotropy of these compounds.
\pacs 71.15.Ap, 71.15.Mb, 71.15.Nc, 71.20.Nr, 65.40.Ba, 65.40.De, 78.20.Ci
\end{abstract}

\section{Introduction}

Polyphosphides are known to form a variety of crystalline structures with different elements \cite{Dolyniuk13, Eschen02,Kraus11}. The electronic structures of polyphosphides can be rationalized by the application of the
Zintl concept \cite{Dell98,Miller11}. Classic Zintl phases usually contain a class of intermetallic compounds that are made up of electro-positive elements (alkali and alkali-earth metals), in which valence electrons are given by electropositive atoms to more electronegative atoms from groups 13 and 15 \cite{Dell98}. These latter can be gained their electron octet by forming chemical bonds and by having pairs of free electrons. In A$_{3}$P$_{7}$, (A = alkali metal) an electron is given from each atom A to allow the formation of the cage P$_{7}^{-3}$ \cite{Dolyniuk13,Manriquez86}, in Ba$_{3}$P%
$_{14}$, three Ba atoms each give two electrons to allow the formation of two
cages  P$_{7}^{-3}$ \cite{Dahlmann73}, in Ba$_{2}$P$_{7}$Cl Zintl salt, the
total charge of two Ba$^{+2}$ cations is compensated by a combination Cl$^{-}
$ of and P$^{-3}$ anions. Juli-Anna Dolyniuk and Kirill Kovnir \cite{Dolyniuk13} have shown that the Ba$_{2}$P$_{7}$Br and Ba$_{2}$P$%
_{7}$I isostructurally include the family of P$_{7}^{-3}$ structures.
A large number of Zint compounds have recently been synthesized and these
structures offer abundant and interesting physical properties, such as
semi-conductivity, superconductivity. In a recent experimental study,
Juli-Anna Dolyniuk and Kirill Kovnir \cite{Dolyniuk13} synthesized the new
Zintl phase Ba$_{2}$P$^{7}$X (barium and indium phosphide) and analyzed its
crystalline structure. According to \cite{Dolyniuk13}, both compounds Ba$_{2}$P%
$_{7}$Br and Ba$_{2}$P$_{7}$I are crystallized in a new type of monoclinic
structure in space group $P2_{1}/m$ (No.~11) and are isostructural to Ba$_{2}$P$_{7}$Cl \cite{Dolyniuk13}. The crystal
structures of Ba$_{2}$P$_{7}$X (X = Cl, Br, I) exhibit the presence of  P$%
_{7}^{-3}$ groups with halogenated anions and barium cations.

Authors think that neither theoretical nor experimental studies of the elastic properties had been carried out. Therefore, such calculations are made in the present work with the inclusion of pressure effects. The results reported in
this paper may be useful for evaluating the potential technological
applications of Ba$_{2}$P$_{7}$X. Knowledge of the elastic constants of
crystalline materials is essential to understand many of their basic
physical properties. In particular, these constants provide information on
the stability and stiffness of the material against externally applied
stresses \cite{Sin'ko08}. Knowledge of the pressure dependence of elastic
constants and lattice parameters is significant for many modern technologies~\cite{Sin'ko08,Zhijiao11}.

\section{Computational details}

Currently, there are different theoretical calculation codes with different
approximations. In our calculations, we use the code CASTEP (CambridgeSerial
Total Energy Package) \cite{Clark05} which is a direct application of the
calculation. All calculations were performed using pseudo-potential plane
wave \textsl{ab~initio} (PP-PW) method based on the density function (DFT). To
determine the structural parameters and elastic moduli of the considered
compounds, there was used a new version of the generalized gradient approximation (GGA),
namely the GGA-PBEsol \cite{Perdew08}, which has been developed specifically
to improve the description of the exchange-correlation in solids.
In all electronic total energy calculations, an ultra-soft Vanderbilt
pseudo-potential \cite{Vanderbilt92} was used to treat the potential seen by
the valence electrons due to the nucleus and electrons of the frozen
nucleus. Ba 5s$^{2}$ 5p$^{6}$ 6s$^{2}$, P 3s$^{2}$ 3p$^{3}$ and X (Cl 3s$^{2}$
3p$^{5}$, Br 4s$^{2}$ 4p$^{5}$, I 5s$^{2}$ 5p$^{6}$ 6s$^{2}$) have been
explicitly treated as valence electron states. Valence's electronic wave
functions were extended into a set of truncated plane wave bases at maximum
plane energy (cutoff energy) of 380~eV. The Brillouin zone (BZ) was sampled
on a 3$\times$5$\times$5 Monkhorst-Pack special $k$ mesh \cite{Monkhorst76}.
For $k$-points were chosen, after a convergence test, in order to ensure sufficiently
accurate calculations.

The fully optimized geometry was carried out with the herein mentioned
convergence criteria: (i)~the difference of total energy between two
consecutive iterations was smaller than 7.57$\times $10$^{-7}$~eV/atom, (ii)
maximum force on any atom was smaller than 0.015~eV/\AA , (iii) stress was
smaller than 0.04~GPa and (iv) atomic displacement was smaller than 0.002 
\AA. The single-crystal elastic constants $C_{ijs}$ were determined via a
linear fitting of the stress-strain curves obtained from first-principles
calculations \cite{Clark05}. The elastic constants were done following the
convergence of these criteria: 5.37$\times 10^{-7}$ eV/atom for total energy,
0.0084~eV/\AA\ for Hellman-Feynman force and 3.54$\times $10$^{-6}$~\AA\ for
maximal ionic displacement. The polycrystalline aggregate elastic moduli,
namely the bulk modulus $B$ and shear modulus $G$, were evaluated via the
Voigt-Reuss-Hill approximations \cite{Voigt28,Hill52}.

\section{Results and discussions}

\subsection{Structural properties}

\begin{figure}[!t]
\centerline{\includegraphics[width=0.95\textwidth]{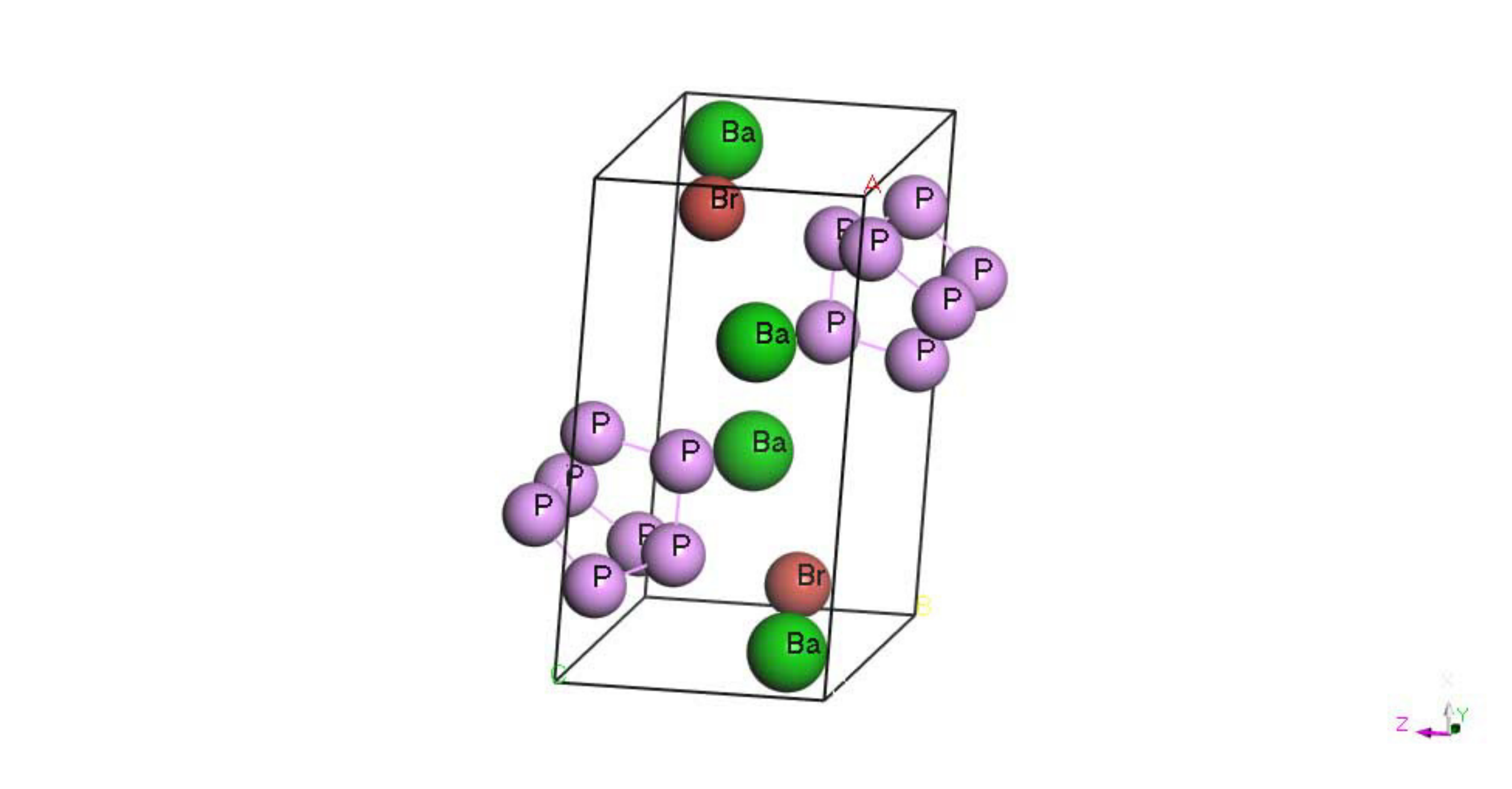}}
\caption{(Colour online) The unit-cell crystalline structure of the monoclinic Zintl phase
Ba$_2$P$_7$Br compound.} \label{fig-smp1}
\end{figure}

The ternary semiconductor compounds Ba$_{2}$P$_{7}$X where (X= Cl, Br, I)
have a monoclinic structure and belongs to the $P2_{1}/m$ (No.~11), with tow inequivalent atomic positions for the barium Ba$_{1}$ and Ba$%
_{2}$, and five inequivalent atomic positions for the phosphorous atoms P$%
_{1}$, P$_{2}$, P$_{3}$, P$_{4}$ and P$_{5}$. One conventional cell of the Ba%
$_{2}$P$_{7}$Cl crystal is depicted in figure~\ref{fig-smp1}.

The conventional cell of Ba$_{2}$P$_{7}$X (X= Cl, Br, I) contains 20 atoms,
4Ba, 14P and 2X(X= Cl, Br, I), which means that the unit cell of Ba$_{2}$P$%
_{7}$X contains two unit formulas (Z=2). The atomic positions are Ba$_{1}$:
2e(X$_{\text {Ba}_{1}}$, 0.25, Z$_{\text {Ba}_{1}}$), Ba$_{2}$: 2e(X$_{\text {Ba}_{2}}$, 0.25, Z$_{\text {Ba}_{2}}$%
), P$_{1}$: 2e(X$_{\text P_{1}}$, 0.25, Z$_{\text P_{1}}$), P$_{2}$: 2e(X$_{P_{2}}$,
0.25, Z$_{\text P_{2}}$), P$_{3}$:~2e(X$_{\text P_{3}}$, 0.25, Z$_{\text P_{3}}$) , P$_{4}$:
4f(X$_{\text P_{4}}$, Y$_{\text P_{4}}$, Z$_{\text P_{4}}$), P$_{5}$: 4f(X$_{\text P_{5}}$, Y$%
_{\text P_{5}}$, Z$_{\text P_{5}}$) and X: 2e(X$_\text{X}$ , 0,25, Z$_\text{X}$). The atoms are
indexed in order to distinguish between the inequivalent crystallographic
positions of the same chemical element. Thus, the crystalline structures of
the title compounds are characterized by 22 parameters not fixed by the
group symmetry, 18 atomic coordinates and three lattice parameters constants
(a, b and c), one angle~$\beta $.

We have used the experiment lattice constant to start calculating the
parameters of the lattice (a, b and c), the angle $\beta $ and the internal
atomic coordinates. The calculated equilibrium crystal parameters for Ba$%
_{2} $P$_{7}$X are presented in table~\ref{tableONE} and table~\ref{tableSECOND} along with the available
experimental data \cite{Schnering81} for the sake of comparison. One can
observe an excellent agreement between the calculated and experimental
values of the parameters of the lattice (a, b and c) and the angle $\beta $,
the maximum relative difference between the calculated values and their
corresponding measured values is very small. In addition, the calculated and the
measured atomic internal coordinates of all atoms of the unit cell match
each other well. This excellent matching serves as a proof of the reliability
and accuracy of the chosen calculation method and provides confidence in the
results of the following calculations of the structural and elastic
properties of the considered system.

We note that the unit cell volume of Ba$_{2}$P$_{7}$I is wider than that of
Ba$_{2}$P$_{7}$Br which is larger than that of Ba$_{2}$P$_{7}$Cl, which can
be attributed to the fact that the I atom radius is larger than that Br than
that greater than that of Cl atom, i.e., the unit-cell volume of ternary
compounds of Zintl type increases when moving down in column VII of the
periodic table.

In order to objectively study the chemical and structural stability of the
monoclinic ternary Ba$_{2}$P$_{7}$X, the cohesive energy $E_\text{coh}$ and formation
enthalpy $\Delta $H are calculated using the following expressions \cite{Wu10}%
:
\begin{equation}
E_\text{coh}=\frac{1}{N_\text{Ba}+N_\text{P}+N_\text{X}}\left[ E_\text{Tot}^{\text {Ba}_{2} \text{P}_{7}\text{X}}-\left(
N_\text{Ba}E_\text{Tot}^{\text{Ba}(\text{atom}) }+N_\text{P}E_\text{Tot}^{\text P(\text{atom})
}+N_\text{X}E_\text{Tot}^{\text X(\text{atom}) }\right) \right] .  \label{1}
\end{equation}

Where these quantities $E_\text{Tot}^{\text {Ba}_{2} \text{P}_{7}\text{X}}$, $E_\text{Tot}^{\text{Ba}(\text{atom}) }$, $E_\text{Tot}^{\text P(\text{atom})
}$ and $E_\text{Tot}^{\text X(\text{atom}) }$ represent the total energy of the primitive cell of Ba$_{2}$P$%
_{7}$X and the total energies of the isolated Ba, P and X atoms,
respectively. N$_\text{Ba}$, N$_\text{P}$ and N$_\text{X}$ are the number of Ba, P and X
atoms in the primitive cell, respectively. The energy of the free atom was
calculated using a cubic box with a large lattice constant that contained
the considered atom. The formation enthalpy $\Delta H$ of Ba$_2$P$_7$X was
calculated using the following expression \cite{Wu10}:
\begin{equation}
\Delta H=\frac{1}{N_\text{Ba}+N_\text{P}+N_\text{X}}\left[ E_\text{Tot}^{\text {Ba}_{2} \text P_{7} \text X}-\left(
N_\text{Ba}E_\text{Tot}^{\text {Ba} (\text {solid}) }+N_\text{P}E_\text{Tot}^{\text P (\text {solid})
}+N_\text{X}E_\text{Tot}^{\text X (\text {solid}) }\right) \right].  \label{2}
\end{equation}

Here, $E_\text{Tot}^{\text {Ba} (\text {solid}) }$, $E_\text{Tot}^{\text {P} (\text {solid}) }$%
and $E_\text{Tot}^{\text {X} (\text {solid}) }$ denote the total energies per atom of
the solid states of the pure elements Ba, P and X, respectively. The
thermodynamic and chemical stabilities of the Ba$_{2}$P$_{7}$Cl, Ba$_{2}$P$%
_{7}$Br and Ba$_{2}$P$_{7}$I compounds can be judged from their formation
enthalpies and cohesive energies. As can be seen from table~\ref{tableONE}, three
considered compounds monoclinic Zintl phase have negative cohesive energies
and formation enthalpies, indicating that they are energetically stable.

\begin{table}[!t]%
\caption{The calculated equilibrium crystal lattice constants (a, b and c, in \AA), angle $\beta$ (in deg), unit-cell volume ($V$, in \AA$^3$), cohesive energy ($E_\text{coh}$, in eV/atom) and formation enthalpy ($\Delta H$, in eV/atom)
 for the monoclinic Zintl phase Ba$_2$P$_7$X (X=Cl, Br, I) compared with the available experimental data.}%
\vspace{2ex}
\small
\begin{center}
\begin{tabular}[b]{|c|c|c|c|c|c|c|}
\hline
Structural & \multicolumn{2}{|c|}{Ba$_{2}$P$_{7}$Cl} & \multicolumn{2}{|c|}{Ba$%
_{2}$P$_{7}$Br} & \multicolumn{2}{|c|}{Ba$_{2}$P$_{7}$I} \\ 
\cline{2-7} parameter & \multicolumn{1}{|c|}{Present} & \multicolumn{1}{|c|}{Expt
 \cite{Schnering81}} & \multicolumn{1}{|c|}{Present} & 
\multicolumn{1}{|c|}{Expt \cite{Schnering81}} & \multicolumn{1}{|c|}{Present} & Expt \cite{Schnering81} \\ 
&work&&work&&work&
\\ \hline
a & 11.618 & 11 .726 & 11.786 & 11.850 & 11.975 & 12.0392 \\ 
b & 6.731 & 6 .829 & 6.779 & 6.835 & 6.834 & 6.8990 \\ 
c & 6.270 & 6.337 & 6.232 & 6.294 & 6.292 & 6.3538 \\ 
$\beta $ & 95.350 & 95.270 & 95.951 & 95.819 & 96.127 & 95.915 \\ 
$V$ & 488.223 & 505.302 & 495.309 & 507.2 & 6.292 & 524.93 \\ 
$E_\text{coh}$ & --5.53 & -- & --5.48 & -- & 512.023 & -- \\ 
$\Delta H$ & --0.38 & -- & --0.31 & -- & --0.280 & -- \\ \hline
\end{tabular}%
\end{center}
	\label{tableONE}
\end{table}%
\vspace{-4mm}
\begin{table}[!t]%
\caption{Optimized atomic coordinates for the monoclinic Zintl
Phase Ba$_2$P$_7$X (X=Cl, Br, I)  in comparison with experiment.}%
\scriptsize
\renewcommand{\arraystretch}{1.3}
\begin{center}
\begin{tabular}[b]{cc|c|c|c|c|c|c|c|c|c|}
\cline{3-11}
&  & \multicolumn{3}{|c|}{Ba$_{2}$P$_{7}$Cl} & 
\multicolumn{3}{|c|}{Ba$_{2}$P$_{7}$Br} & \multicolumn{3}{|c|}{%
Ba$_{2}$P$_{7}$I} \\ \cline{3-11}
&  & { $x$} & { $y$} & { $z$} & { $x$} & 
{ $y$} & { $z$} & { $x$} & { $y$} & 
{ $z$} \\ \hline
\multicolumn{1}{|c}{{ Ba}$_{1}$} & \multicolumn{1}{|c|}%
{ present} & { 0.0466} & { 0.25} & 
{ 0.229} & { 0.0459} & { 0.25} & 
{ 0.2113} & { 0.05} & { 0.25} & 
{ 0.1968} \\ \cline{2-11}
\multicolumn{1}{|c}{ (2e)} & \multicolumn{1}{|c|}{{ %
Expt \cite{Schnering81}}} & { 0.045} & { 0.25} & 
{ 0.2292} & { --} & { --} & { --} & 
{ --} & { --} & { --} \\ \hline
\multicolumn{1}{|c}{{ Ba}$_{1}$} & \multicolumn{1}{|c|}%
{ Present} & { 0.6524} & { 0.25} & 
{ 0.4348} & { 0.6469} & { 0.25} & 
{ 0.4343} & { 0.638} & { 0.25} & 
{ 0.4291} \\ \cline{2-11}
\multicolumn{1}{|c}{ (2e)} & \multicolumn{1}{|c|}{{ %
Expt \cite{Schnering81}}} & { 6524} & { 0.25} & 
{ 0.4332} & { --} & { --} & { --} & 
{ --} & { --} & { --} \\ \hline
\multicolumn{1}{|c}{ X} & \multicolumn{1}{|c|}{ Present%
} & { 0.9115} & { 0.25} & { 0.6379} & 
{ 0.9145} & { 0.25} & { 0.6558} & 
{ 0.9142} & { 0.25} & { 0.001} \\ 
\cline{2-11}
\multicolumn{1}{|c}{ (2e)} & \multicolumn{1}{|c|}{{ %
Expt \cite{Schnering81}}} & { 0.9113} & { 0.25} & 
{ 0.6366} & { --} & { --} & { --} & 
{ --} & { --} & { --} \\ \hline
\multicolumn{1}{|c}{{ P}$_{1}$} & \multicolumn{1}{|c|}%
{ Present} & { 0.452} & { 0.25} & 
{ 0.0098} & { 0.4501} & { 0.25} & 
{ 0.0067} & { 0.4481} & { 0.25} & 
{ 0.6617} \\ \cline{2-11}
\multicolumn{1}{|c}{ (2e)} & \multicolumn{1}{|c|}{{ %
Expt \cite{Schnering81}}} & { 0.451} & { 0.25} & 
{ 0.01} & { --} & { --} & { --} & 
{ --} & { --} & { --} \\ \hline
\multicolumn{1}{|c}{{ P}$_{2}$} & \multicolumn{1}{|c|}%
{ Present} & { 0.2212} & { 0.25} & 
{ 0.6764} & { 0.222} & { 0.25} & 
{ 0.6661} & { 0.2234} & { 0.25} & 
{ 0.6609} \\ \cline{2-11}
\multicolumn{1}{|c}{ (2e)} & \multicolumn{1}{|c|}{{ %
Expt \cite{Schnering81}}} & { 0.2222} & { 0.25} & 
{ 0.6767} & { --} & { --} & { --} & 
{ --} & { --} & { --} \\ \hline
\multicolumn{1}{|c}{{ P}$_{3}$} & \multicolumn{1}{|c|}%
{ Present} & { 0.4076} & { 0.25} & 
{ 0.6696} & { 0.4063} & { 0.25} & 
{ 0.6639} & { 0.405} & { 0.25} & 
{ 0.6609} \\ \cline{2-11}
\multicolumn{1}{|c}{ (2e)} & \multicolumn{1}{|c|}{{ %
Expt \cite{Schnering81}}} & { 0.408} & { 0.25} & 
{ 0.6722} & { --} & { --} & { --} & 
{ --} & { --} & { --} \\ \hline
\multicolumn{1}{|c}{{ P}$_{4}$} & \multicolumn{1}{|c|}%
{ Present} & { 0.1939} & { 0.4989} & 
{ 0.8834} & { 0.1959} & { 0.4980} & 
{ 0.8763} & { 0.1986} & { 0.0016} & 
{ 0.8662} \\ \cline{2-11}
\multicolumn{1}{|c}{ (2e)} & \multicolumn{1}{|c|}{{ %
Expt \cite{Schnering81}}} & { 0.1946} & { 0.0047} & 
{ 0.8828} & { --} & { --} & { --} & 
{ --} & { --} & { --} \\ \hline
\multicolumn{1}{|c}{{ P}$_{5}$} & \multicolumn{1}{|c|}%
{ Present} & { 0.3142} & { 0.0768} & 
{ 0.1483} & { 0.3139} & { 0.0784} & 
{ 0.1441} & { 0.3141} & { 0.0797} & 
{ 0.1370} \\ \cline{2-11}
\multicolumn{1}{|c}{ (2e)} & \multicolumn{1}{|c|}{{ %
Expt \cite{Schnering81}}} & { 0.3141} & { 0.0789} & 
{ 0.1466} & { --} & { --} & { --} & 
{ --} & { --} & { --} \\ \hline
\end{tabular}%
\end{center}
\label{tableSECOND}
\end{table}%

\begin{figure}[!t]
	\includegraphics[width=0.295\textwidth]{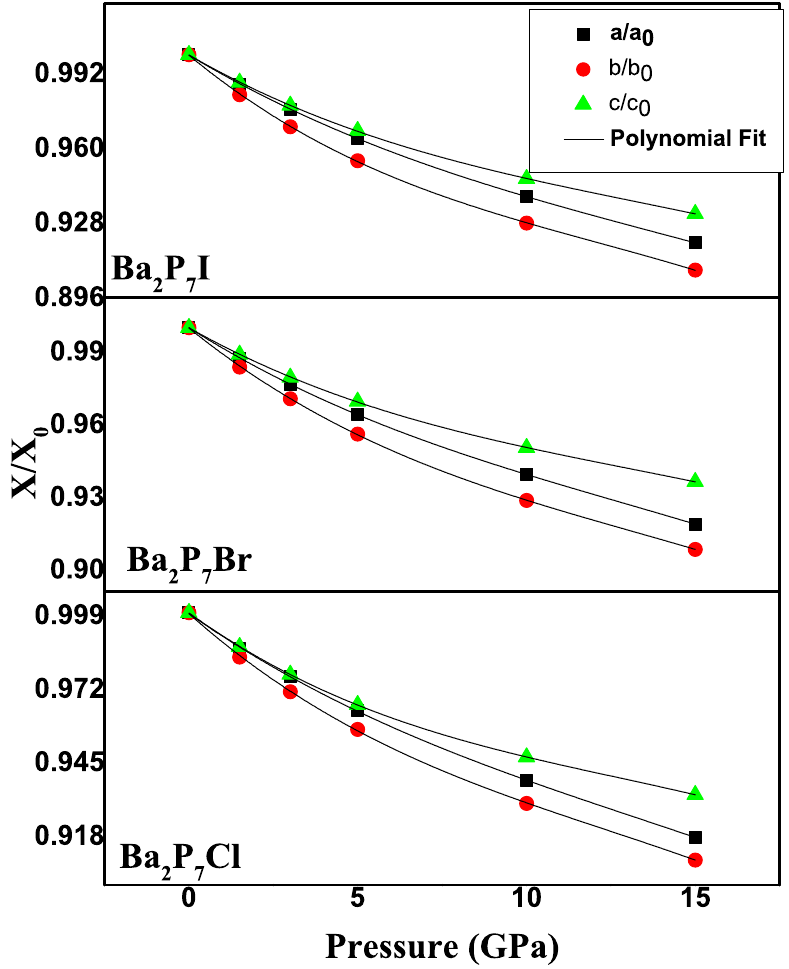}
	\includegraphics[width=0.3\textwidth]{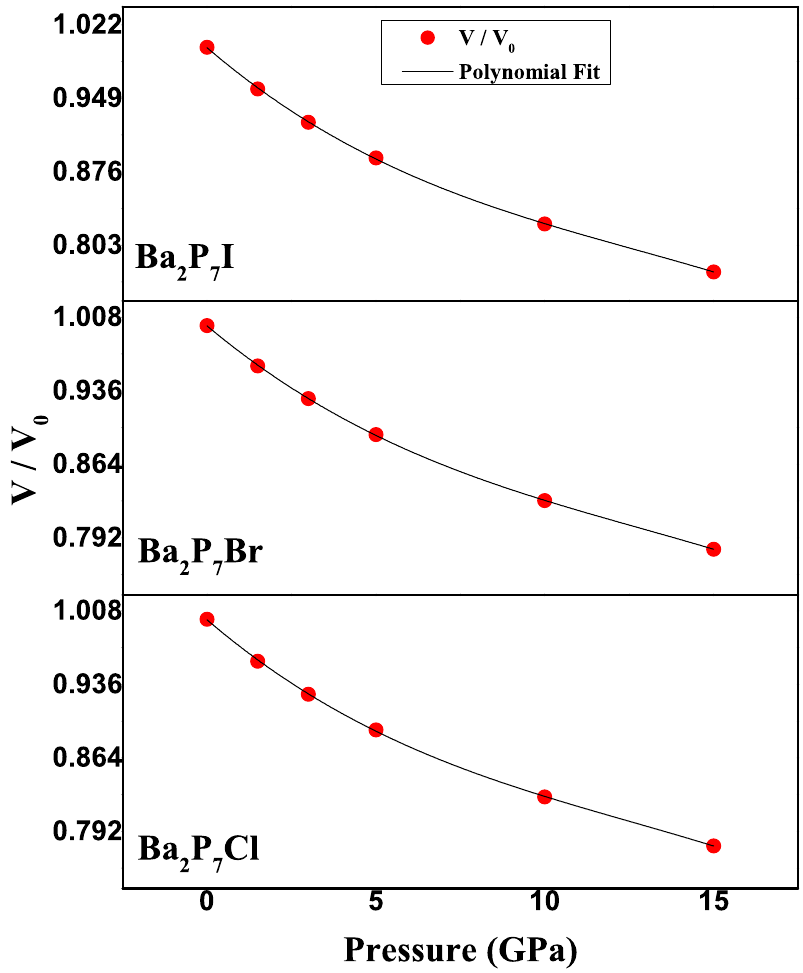}
	\includegraphics[width=0.305\textwidth]{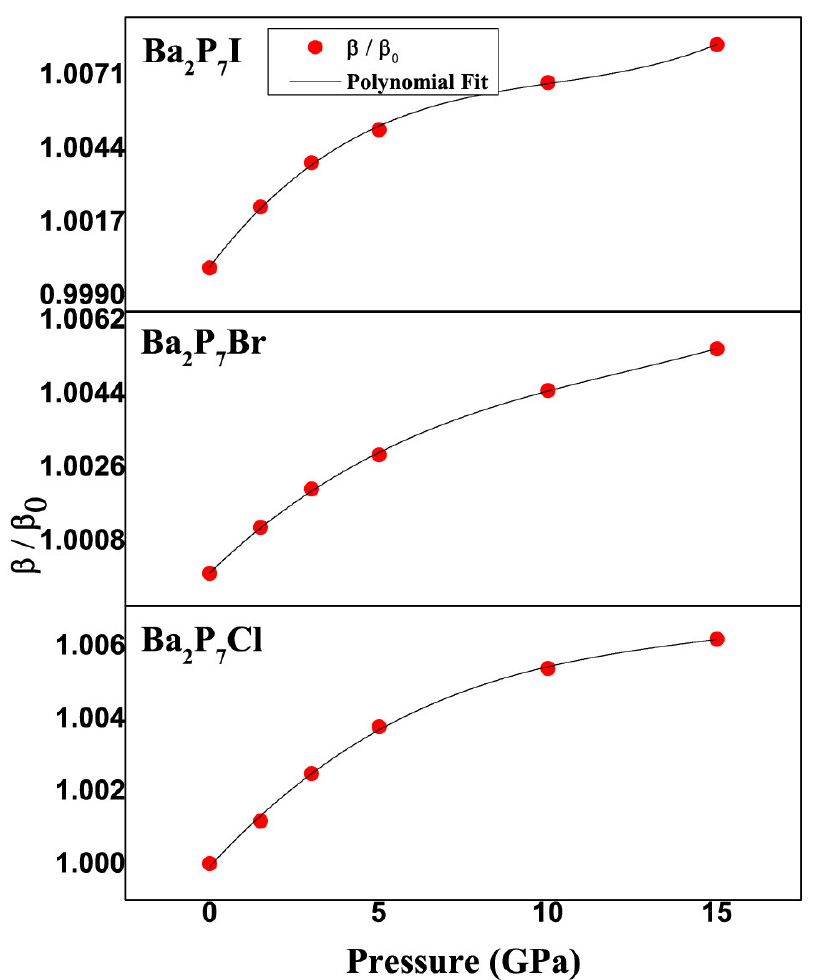}
	\caption{(Colour online) Pressure-dependent variations of the lattice constants a, b and c,
		the unit-cell volume and the angle $\beta$ for the monoclinic Zintl phase
		Ba$_2$P$_7$X (X=Cl, Br, I). The ``0'' subscript denotes the value of the
		parameter at zero pressure.} \label{fig-smp2}
\end{figure}

The most frequently used methods of testing the reliability of the obtained theoretical
results consist in comparing the numerical values of a property
obtained by different theoretical procedures. For this problem, the bulk
module $B$ was used as a test parameter. The calculations of the unit-cell
volume $V$ and the total energy $E_\text{Tot}$ of a solid for different values of
the pressure $P$ provide a convenient method for estimating the bulk modulus $B$
and its pressure derivative $B'$. For this purpose, the structural parameters
of the test compounds were calculated at fixed applied hydrostatic pressures
in the range of 0 to 15 GPa with a pitch of 5 GPa, such an option being
implemented in the CASTEP code and makes it possible to find an optimized
structure at any axial or hydrostatic pressure, so that the hydrostatic
pressure can significantly affect the physical properties of the materials.
One of the most obvious manifestations of the effect of the application of
hydrostatic pressure to a material is a decrease of its volume and lattice
constants. Figure~\ref{fig-smp2} demonstrates the dependence in pressure of the normalized
lattice constants (a/a$_{0}$, b/b$_{0}$ and c/c$_{0}$), the normalized
unit-cell volume $(V/V_{0})$ and the normalized angle $\beta $ ($\beta $/$%
\beta _{0}$). We have fitted these quantities a/a$_{0}$, b/b$_{0}$, c/c$_{0}$
and $(V/V_{0})$ (where a$_{0}$, b$_{0}$, c$_{0}$ and $V_{0}$ are the lattice
parameters and unit-cell volume at zero pressure) using a polynomial
expression in the following form:
\begin{equation}
\frac{X\left( P\right) }{X_{0}}=1+B_\text{X}P+\sum_{n=2}^{3}K_{n}P^{n},  \label{3}
\end{equation}
where X = a, b, c, $V$. The obtained linear compressibilities B$_\text{a}$, B$_\text{b}$
and B$_\text{c}$ of the lattice parameters a, b and c, respectively, and the
volume compressibility B$_{V}$ were used to estimate the bulk modulus B as
follows:
\begin{equation}
B=\frac{1}{B_\text{a}+B_\text{b}+B_\text{c}}\,,  \label{4}
\end{equation}
\begin{equation}
B=\frac{1}{B_{V}}.  \label{5}
\end{equation}

We observe a third-order polynomial dependence in all curves as the pressure
increases from 0 GPa to 15 GPa. Cellular axes and volume decrease with
pressure while ${B}/{B}_{0}$ increases. Moreover, we can see that the ratio $\text{b}/\text{b}_{0}$ decreases more rapidly than $\text{a}/\text{a}_{0}$ and $\text{c}/\text{c}_{0}$ in the
three materials, respectively, which indicates that the axis b is much more
compressible than the axes a and c. The deduced values of compressibility
$B\text{x}=-\frac{1}{x}\frac{\rd x}{\rd P}$. From the fit, for the three materials Ba$%
_{2}$P$_{7}$X (X = Cl, Br, I), respectively. The following relations were
obtained from these calculations:

for the Ba$_{2}$P$_{7}$Cl:

$B_\text{a}$ = 0.00885 GPa$^{-1}$, $B_\text{b}$ = 0.01102 GPa$^{-1}$, $B_\text{c}$ =
0.00888 GPa$^{-1}$ and $B_{V}$ = 0.02831 GPa$^{-1}$. The obtained
polynomials the fit are:
\begin{eqnarray}
\left\{
\begin{array}{lll}
\frac{\text a}{\text a_{0}}&=&1-0.00885P+3.38214\times 10^{-4}P^{2}-8.45661\times 10^{-6}P^{3}, \\ 
\frac{\text b}{\text b_{0}}&=&1-0.01102P+5.57587\times 10^{-4}P^{2}-1.5016\times 10^{-5}P^{3}, \\ 
\frac{\text c}{\text c_{0}}&=&1-0.00888P+4.88156\times 10^{-4}P^{2}-1.28474\times 10^{-5}P^{3}, \\ 
\frac{V}{V_{0}}&=&1-0.02831P+1.15\times 10^{-3}P^{2}-3.97585\times 10^{-5}P^{3}, \\ 
\frac{\beta }{\beta _{0}}&=&1+1.02\times 10^{-3}P-6.0406\times 10^{-5}P^{2}+1.34967\times 10^{-6}P^{3}\,,
\end{array}%
\right.  \label{6}
\end{eqnarray}
for the Ba$_{2}$P$_{7}$Br:
$B_\text{a}$ = 0.0088 GPa$^{-1}$, $B_\text{b}$ = 0.01122 GPa$^{-1}$, $B_\text{c}$ = 0.00782
GPa$^{-1}$ et $B_{V}$ = 0.02765 GPa$^{-1}$. The obtained polynomials the fit
are:
\begin{eqnarray}
\left\{ 
\begin{array}{lll}
\frac{\text a}{\text a_{0}}&=&1-0.0088P+3.704\times 10^{-4}P^{2}-9.6174\times 10^{-6}P^{3}, \\ 
\frac{\text b}{\text b_{0}}&=&1-0.01122P+5.5058\times 10^{-4}P^{2}-1.39338\times 10^{-5}P^{3}, \\ 
\frac{\text c}{\text c_{0}}&=&1-0.00782P+3.86272\times 10^{-4}P^{2}-9.86085\times 10^{-5}P^{3}, \\ 
\frac{V}{V_{0}}&=&1-0.02765P+1.43\times 10^{-3}P^{2}-3.70713\times 10^{-5}P^{3}, \\ 
\frac{\beta }{\beta _{0}}&=&1+7.95806\times 10^{-4}P-4.81896\times 10^{-5}P^{2}+1.29935\times 10^{-6}P^{3},%
\end{array}%
\right.  \label{7}
\end{eqnarray}
for the Ba$_{2}$P$_{7}$I:
$B_\text{a}$ = 0.00865 GPa$^{-1}$, $B_\text{b}$ = 0.0119 GPa$^{-1}$, $B_\text{c}$ = 0.00829
GPa$^{-1}$ and $B_{V}$ = 0.02879 GPa$^{-1}$. The obtained polynomials the fit
are:
\begin{eqnarray}
\left\{ 
\begin{array}{lll}
\frac{\text a}{\text a_{0}}&=&1-0.00865P+3.38005\times 10^{-4}P^{2}-7.91834\times 10^{-6}P^{3}, \\ 
\frac{\text b}{\text b_{0}}&=&1-0.01196P+6.6066\times 10^{-4}P^{2}-1.81681\times 10^{-5}P^{3}, \\ 
\frac{\text c}{\text c_{0}}&=&1-0.00829P+4.01591\times 10^{-4}P^{2}-1.00773\times 10^{-5}P^{3}, \\ 
\frac{V}{V_{0}}&=&1-0.02879P+1.53\times 10^{-3}P^{2}-4.02183\times 10^{-5}P^{3}, \\ 
\frac{\beta }{\beta _{0}}&=&1+1.62\times 10^{-3}P-1.41829\times 10^{-4}P^{2}+4.65511\times 10^{-6}P^{3}.%
\end{array}%
\right.  \label{8}
\end{eqnarray}

When the pressure changes from 0 to 15 GPa, a, b and c decrease
approximately 8~\%, 9~\% and 6~\%, respectively, in the three materials Ba$_{2}$P$%
_{7}$X (X=Cl, Br, I). Consequently, the axis b is the most compressible
under external pressure, and the c axis is the least compressible; the
effect of the pressure on the axis b is much greater than on the c axis.
Thus, Ba$_{2}$P$_{7}$X is anisotropic in compressibility.

\begin{figure}[!t]
\includegraphics[width=0.33\textwidth]{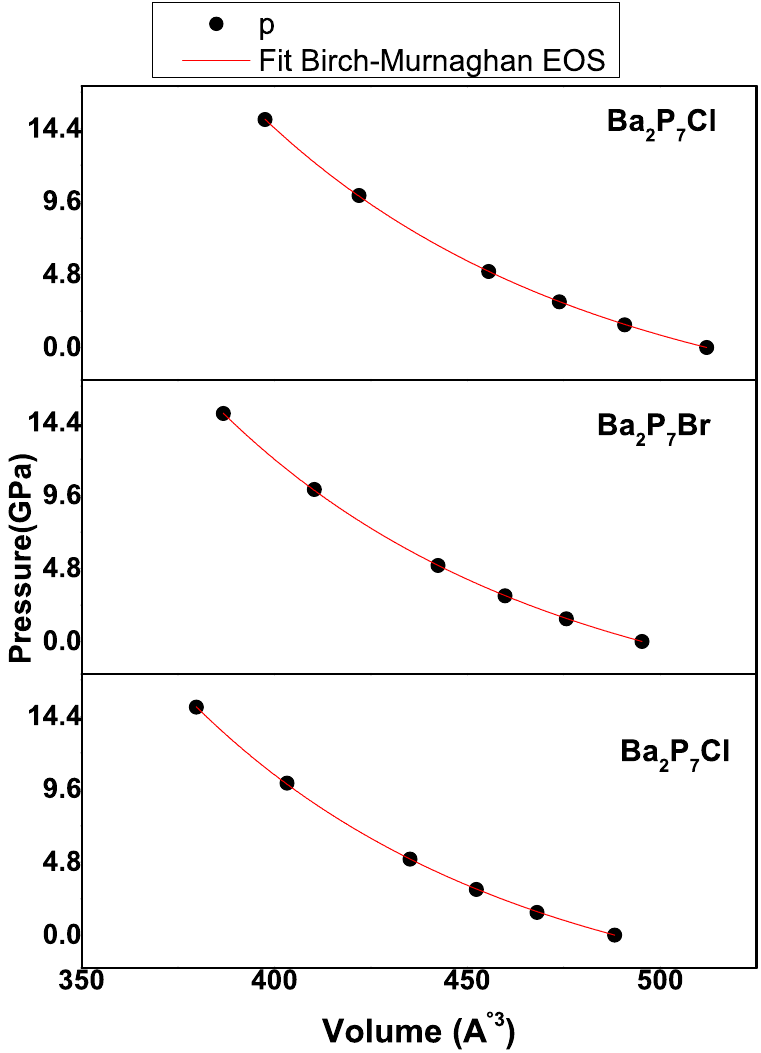}
\includegraphics[width=0.33\textwidth]{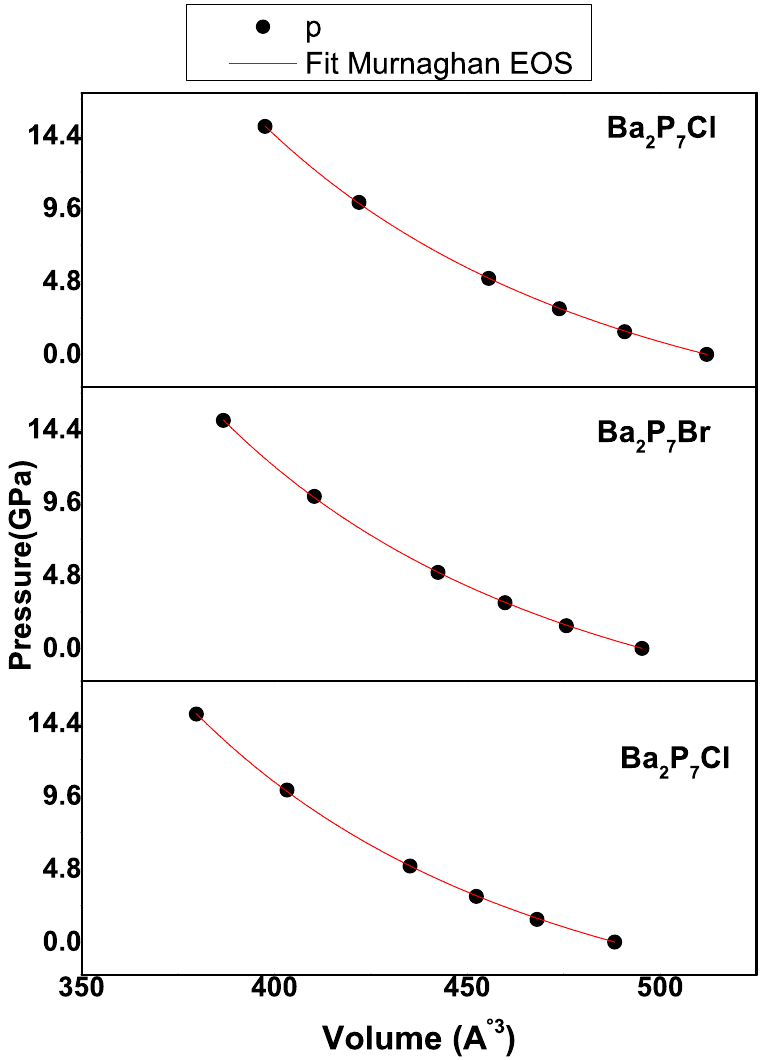}
\includegraphics[width=0.33\textwidth]{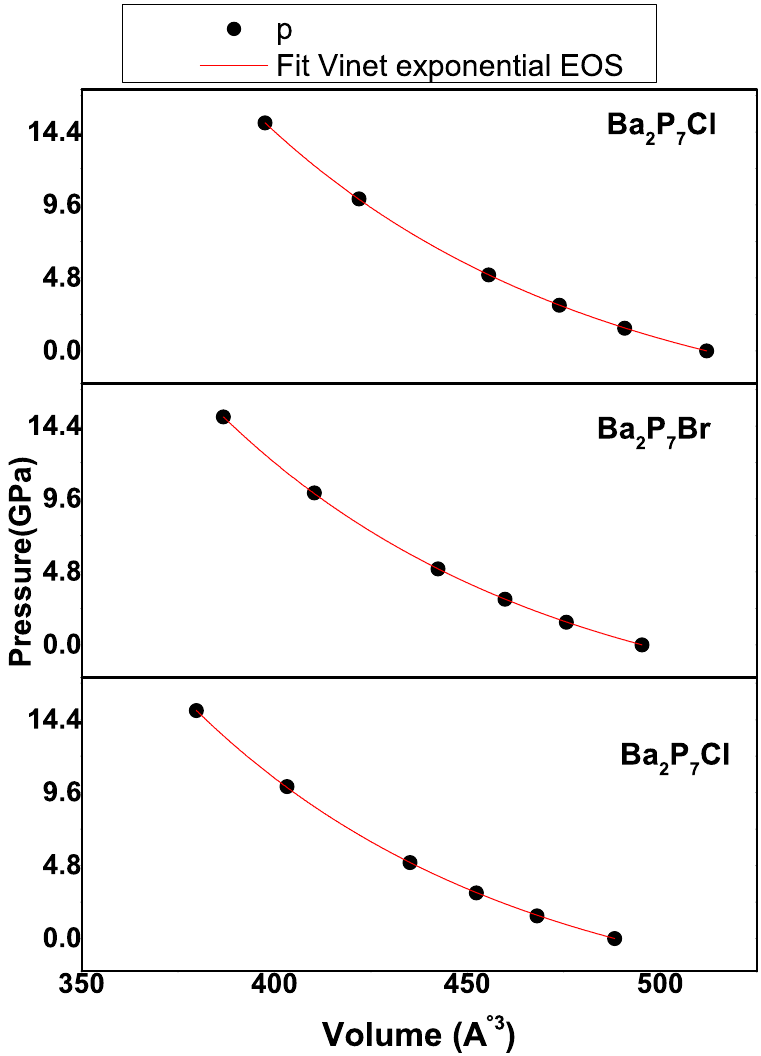}
\centerline{
\includegraphics[width=0.43\textwidth]{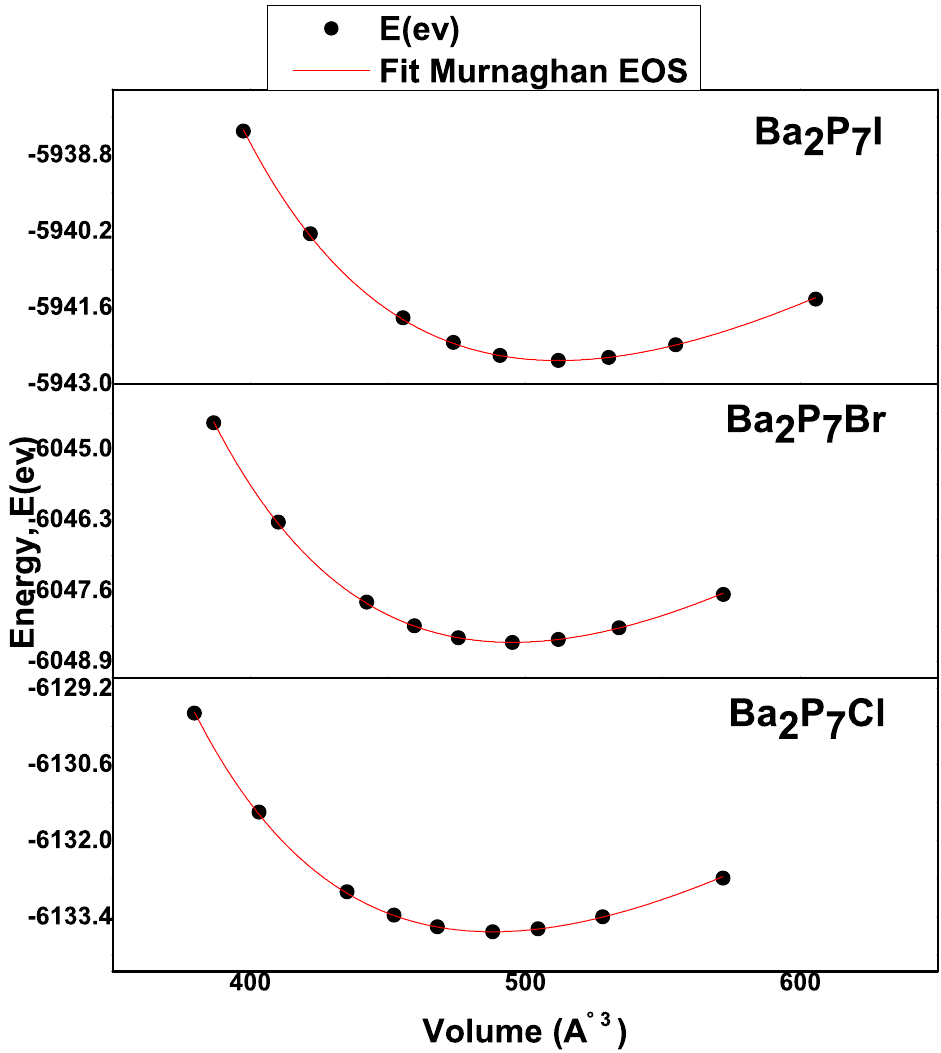}
\includegraphics[width=0.43\textwidth]{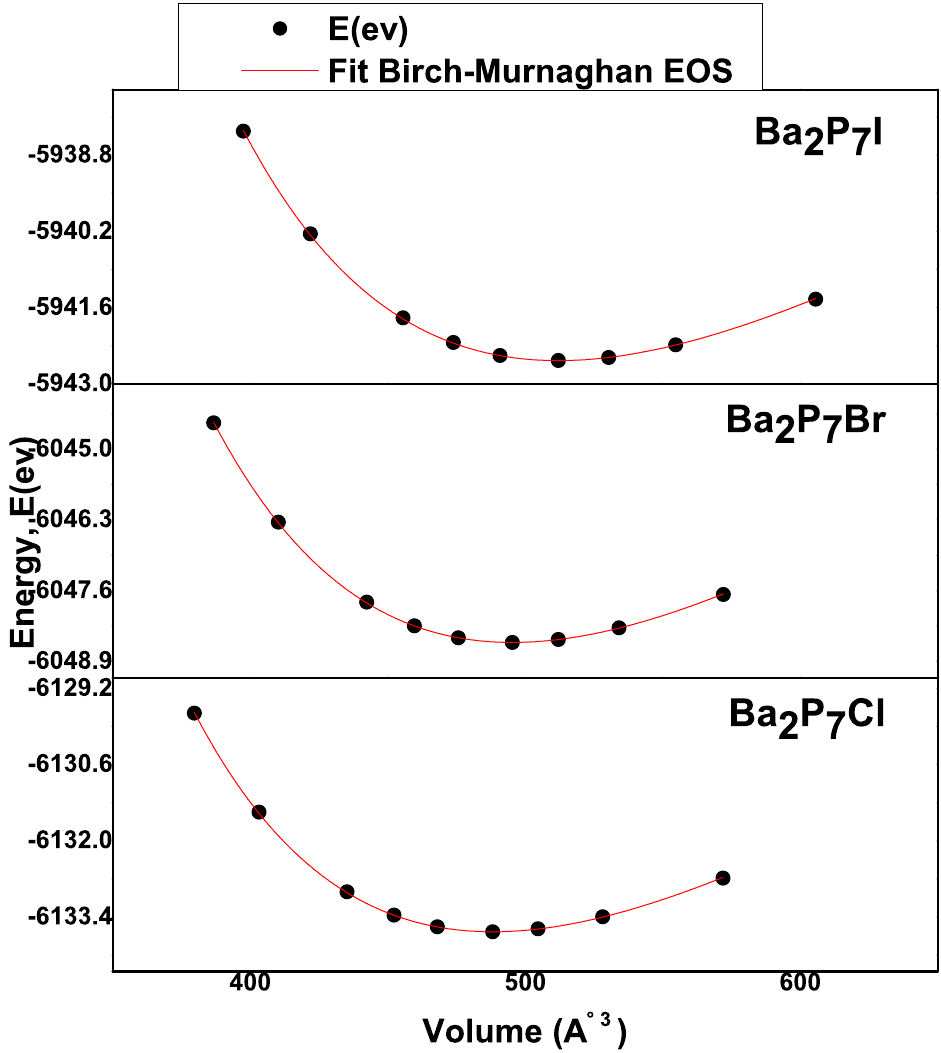}}
\caption{(Colour online) Calculated pressure ($P$) and total energy ($E$) versus unit-cell volume $V$
 and fits to Birch-Murnaghan, Murnaghan and Vinet equation of states for the
 Ba$_2$P$_7$Cl, Ba$_2$P$_7$Br and Ba$_2$P$_7$I compounds.} \label{fig-smp3}
\end{figure}

In the present work, the pressure versus volume ($P-V$) data were fitted to
the Birche Murnaghan \cite{Ambrosch-Draxl06,Hebbache04}, Murnaghan $P-V$ EOS~\cite{Birch78}, and Vinet exponential equations of state \cite{Fu83} equation
of state, and the energy versus volume ($E_\text{Tot}-V$) data were fitted to the
Birche Murnaghan \cite{Murnaghan44} and Murnaghan equations of state \cite%
{Birch47}. Figure~\ref{fig-smp3} shows the obtained results from the bulk modulus $B$ and its
pressure derivative $B'$ is given in table~\ref{table333}. One can appreciate the good
agreement between the values of the bulk modulus $B$ obtained from different
procedures, the linear compressibilities ($B_\text{a}$, $B_\text{b}$ and $B_\text{c}$),
the volume compressibility ($B_{V}$) and the EOSs fits. This constitutes a
good proof for the reliability of our calculations. The obtained bulk
modulus values in our work will be compared to the corresponding ones that
will be achieved from the elastic constants later on.

\begin{table}[!b]%
\caption{Calculated bulk modulus (B, in GPa) and its pressure derivative $B'$.}%
\vspace{2ex}
\renewcommand{\arraystretch}{1.3}
\begin{center}
\begin{tabular}[b]{c|c|c|c|c|c|c|c|c|c|}
\cline{2-10}
& \multicolumn{3}{|c|}{Ba$_{2}$P$_{7}$Cl} & \multicolumn{3}{|c|}{Ba$
_{2}$P$_{7}$Br} & \multicolumn{3}{|c|}{Ba$_{2}$P$_{7}$I} \\ \hline
\multicolumn{1}{|c|}{\small $B$} & {\small 34.013}$^\text{a}$ & {\small 34.014}$%
^\text{b} $ & {\small 32.93}$^\text{c}$ & {\small 34.29}$^\text{a}$ & {\small 34.75}$^\text{b}$
& {\small 33.75}$^\text{c}$ & {\small 32.94}$^\text{a}$ & {\small 35.2}$^\text{b}$ & 
{\small 32.08}$^\text{c}$ \\ \cline{2-10}
\multicolumn{1}{|c|}{} & {\small 33.95}$^\text{d}$ & {\small 33.31}$^\text{e}$ & 
{\small 30.87}$^\text{f}$ & {\small 34.76}$^\text{d}$ & {\small 34.12}$^\text{e}$ & {\small %
32.19}$^\text{f}$ & {\small 33.12}$^\text{d}$ & {\small 32.46}$^\text{e}$ & {\small 30.21}$%
^\text{f}$ \\ \cline{2-10}
\multicolumn{1}{|c|}{} & {\small 31.26}$^\text{g}$ & {\small 34.78}$^\text{h}$ & 
{\small 35.32}$^\text{i}$ & {\small 32.46}$^\text{g}$ & {\small 35.91}$^\text{h}$ & {\small %
36.16}$^\text{i}$ & {\small 30.68}$^\text{g}$ & {\small 34.67}$^\text{h}$ & {\small 34.73}$%
^\text{i}$ \\ \hline
\multicolumn{1}{|c|}{\small $B'$} & {\small 4.89}$^\text{c}$ & {\small 4.14}$^\text{d}$
& {\small 4.61}$^\text{e}$ & {\small 4.89}$^\text{c}$ & {\small 4.15}$^\text{d}$ & {\small %
4.61}$^\text{e}$ & {\small 5.01}$^\text{c}$ & {\small 4.23}$^\text{d}$ & {\small 4.73}$^\text{e}$
\\ \cline{2-10}
\multicolumn{1}{|c|}{} & {\small 5.16}$^\text{f}$ & {\small 5.41}$^\text{g}$ & -- & 
{\small 5.00}$^\text{f}$ & {\small 5.27}$^\text{g}$ & -- & {\small 5.23}$^\text{f}$ & 
{\small 5.47}$^\text{g}$ & -- \\ \hline
\end{tabular}%
\end{center}
	\label{table333}
\end{table}%

a Calculated from GGA PBEsol,

b Calculated from Hill's approximation,

c Calculated from Vinet $P-V$ EOS \cite{Fu83},

d Calculated from Murnaghan $P-V$ EOS \cite{Birch78},

e Calculated from Birch-Murnaghan $P-V$ EOS \cite{Ambrosch-Draxl06},

f Calculated from Murnaghan $E-V$ EOS \cite{Murnaghan44},

g Calculated from Birch-Murnaghan $E-V$ EOS \cite{Birch47},

h Calculated from linear compressibilities $B=1/(B_\text{a}+B_\text{b}+B_\text{c})$,

i Calculated from the compressibility $1/ B_V$.

\subsection{Elastic properties}

\subsubsection{Single-crystal elastic constants}

Elastic constants $C_{ijs}$ of materials are important parameters because
they provide information on their response when a stress is applied to
the material \cite{Westbrook00}.

For monoclinic crystals, there are 13 independent elastic constants, namely,
$C_{11}$, $C_{22}$, $C_{33}$, $C_{44}$, $C_{55}$, $C_{66}$, $C_{12}$, $C%
_{13} $,$C_{15}$, $C_{23}$, $C_{25}$, $C_{35}$ and $C_{46}$. When the $C_{11}$, $C%
_{22}$ and $C_{33}$ represent the stiffness of the material when a uniaxial
stress is applied along the principal $X$, $Y$ and $Z$ axes, respectively. $C_{44}$
measures the shear elastic modulus along $Y$-axis on $Z$-plane; $C_{55}$
measures the shear elastic modulus along $Z$-axis on $X$-plane. Thus, $C_{66}$
represents the shear along $X$-axis on the $Y$-plane. The complete set of the
calculated independent elastic constants $C_{ijs}$ of the Ba$_{2}$P$_{7}$Cl,
Ba$_{2}$P$_{7}$Br and Ba$_{2}$P$_{7}$I compounds. The present work is the
first attempt to calculate the elastic constants $C_{ijs}$ of the title
compounds. No experimental or theoretical values for these quantities are
reported in the literature, which is why comparison with other results is
not possible. From the obtained results, we can make the following
conclusions:

(i) The values $C_{11}$, $C_{22}$ and $C_{33}$ are noticed to be larger than
the ones of $C_{44}$, $C_{55}$, $C_{66}$, $C_{12}$, $C_{13}$, $C_{15}$, $C%
_{25}$, $C_{35}$ and $C_{46}$, which denotes that the considered system is
of a bigger resistance to unidirectional compression than to shear
deformation.

(ii) The stiffness-to-uniaxial stress along the crystallographic a, b and c
axes, respectively, is replicated by $C_{11}$, $C_{22}$ and $C_{33}$ elastic
constants. For three compounds Ba$_{2}$P$_{7}$Cl, Ba$_{2}$P$_{7}$Br and Ba$%
_{2}$P$_{7}$I, the obtained values for elastic constants $C_{11}$, $C_{22}$, C%
$_{33}$ under external pressure of 0 to 15~GPa by 5~GPa step are roughly equal
when 0~GPa is pressed, but with an external pressure variation of 0 to 15~GPa,
we observe an increase of $C_{11}$, $C_{22}$ et $C_{33}$, with \ $C_{22}$\
increase of about 64~\% compared with $C_{33}$, $C_{11}$ values, which is an
increase by less than 64~\% for the three compounds B$_{2}$P$_{7}$X(X=Cl,
Br, I) indicating that the three compounds are relatively more compressible
when compressed along the [010] crystallographic directions than along the
[100] and [001]. These results agree totally with the results already
obtained from the study of the pressure dependence of the lattice parameters.

(iii) To be mechanically stable, the calculated zero-pressure single-crystal
elastic constants $C_{ijs}$ of monoclinic crystals should satisfy the
following stability criteria \cite{Wu07}:
\begin{equation}
\left\{ 
\begin{array}{c}
C_{ii} > 0,\quad i=1,2,3,4,5,6\,, \\ 
\left[ C_{11}+C_{22}+C_{33}+2\left( C_{12}+C_{13}+C_{23}\right) \right]
> 0, \\ 
\big( C_{33}C_{55}-C_{35}^{2}\big) > 0, \quad \big(
C_{44}C_{66}-C_{46}^{2}\big) > 0,\quad \left( C_{22}+C_{33}-2C_{23}\right)
> 0, \\ 
\left[ C_{22}\big( C_{33}C_{55}-C_{35}^{2}\big)
+2C_{23}C_{25}C_{35}-C_{23}^{2}C_{55}-C_{25}^{2}C_{33}\right] > 0, \\ 
\left\{ 
\begin{array}{c}
2\left[ C_{15}C_{25}\left( C_{33}C_{12}-C_{13}C_{23}\right)
+C_{15}C_{35}\left( C_{22}C_{13}-C_{12}C_{23}\right) +C_{25}C_{35}\left(
C_{11}C_{23}-C_{12}C_{13}\right) \right]  \\ 
-\left[ C_{15}^{2}\left( C_{22}C_{33}-C_{23}^{2}\right) +C_{25}^{2}\left(
C_{11}C_{33}-C_{13}^{2}\right) +C_{35}^{2}\left(
C_{11}C_{22}-C_{12}^{2}\right) \right] +C_{55}g%
\end{array}%
\right\} > 0, \\ 
g=C_{11}C_{22}C_{33}-C_{11}C_{23}^{2}-C_{22}C_{13}^{2}-C_{33}C_{12}^{2}+C_{12}C_{13}C_{23}\,.%
\end{array}%
\right.   \label{9}
\end{equation}

Thus, we can assert that the monoclinic Zintl phase Ba$_{2}$P$_{7}$X is in a
mechanically stable state.

(iv) The crystal's mechanical stability at any pressure, which requires
strain energy to be positive, is confirmed when the set of elastic constants
$C_{ijs}$ responds to special restrictions. This criterion is fulfilled if a
symmetric matrix $G_{ij}\ \ $is of a positive determinant \cite{Sin'ko02}.
The symmetric matrix $\tilde{G}_{ij}$ for any structure type is defined as
follows:

\begin{equation}
\tilde{G}=\left\vert 
\begin{array}{cccccc}
\tilde{C}_{11} & \tilde{C}_{12} & \tilde{C}_{13} & 2C_{14} & 2C_{15} & 
2C_{16} \\ 
\tilde{C}_{21} & \tilde{C}_{22} & \tilde{C}_{23} & 2C_{24} & 2C_{25} & 
2C_{26} \\ 
\tilde{C}_{31} & \tilde{C}_{32} & \tilde{C}_{33} & 2C_{34} & 2C_{35} & 
2C_{36} \\ 
2C_{41} & 2C_{42} & 2C_{43} & 4\tilde{C}_{44} & 4C_{45} & 4C_{46} \\ 
2C_{52} & 2C_{52} & 2C_{53} & 4C_{54} & 4\tilde{C}_{55} & 4C_{56} \\ 
2C_{61} & 2C_{62} & 2C_{63} & 4C_{64} & 4C_{65} & 4\tilde{C}_{66}%
\end{array}%
\right\vert .  \label{10}
\end{equation}

Here, $\tilde{C}_{\alpha \alpha }$=$C_{\alpha \alpha }-P$, where $\alpha $=1,
2, \ldots, 6, and $\tilde{C}_{12}=C_{12}+P$ , $\tilde{C}%
_{13}=C_{13}+P$, $\tilde{C}_{23}=C_{23}+P$. The calculated values for
the $C_{ijs}$ of the monoclinic Ba$_{2}$P$_{7}$X (X= Cl, Br, I) compounds in the
pressure range of 0--15~GPa obey these conditions well, which means that this
compound remains mechanically stable in the considered pressure range.

\begin{figure}[!t]
\includegraphics[width=0.49\textwidth]{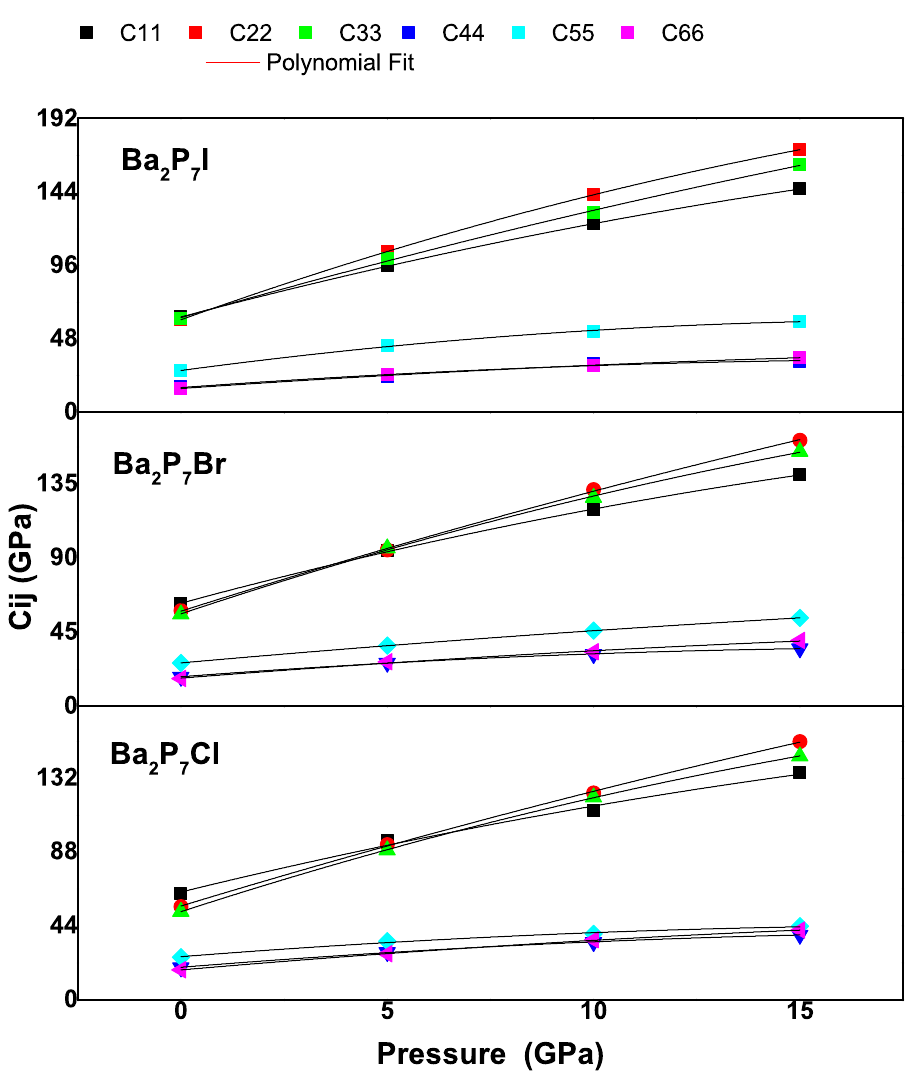}
\includegraphics[width=0.49\textwidth]{m1}
\caption{(Colour online) The calculated pressure dependence of the independent elastic constants $C_{ijs}$ of the monoclinic
 Zintl phase Ba$_2$P$_7$X. The symbols indicate the calculated results.
 The lines represent the results of fitting these theoretical results to a second-order polynomial.} \label{fig-smp4}
\end{figure}

(v) Figure~\ref{fig-smp4} shows the pressure dependence of the 13 independent elastic
constants of the monoclinic compounds for pressures up to 15~GPa. Apart from $C%
_{25}$ and $C_{46}$, the remainder of the elastic constants $C_{ij}$
increase monotonously with an increasing pressure but with different
sensitivities. $C_{25}$ and $C_{46}$ decrease monotonously with an increasing
pressure. The lines represent the second-order polynomial fits to the results.
The fit results given by the following expressions for the three
compounds Ba$_{2}$P$_{7}$Cl, Ba$_{2}$P$_{7}$Br and Ba$_{2}$P$_{7}$I,
respectively, are as follows:
\begin{eqnarray}
&&\left\{ 
\begin{array}{c}
C_{11}=64.25+6.004P-0.088P^{2}, \\ 
C_{22}=55.89+7.47P-0.065P^{2}, \\ 
C_{33}=52.53+7.98P-0.12P^{2}, \\ 
C_{44}=19.44+1.97P-0.046P^{2}, \\ 
C_{55}=25.77+1.91P-0.048P^{2}, \\ 
C_{66}=17.86+2.12P-0.036P^{2},%
\end{array}%
\right. 
\left\{ 
\begin{array}{c}
C_{12}=18.76+2.53P-0.017P^{2}, \\ 
C_{13}=28.32+5.17P-0.075P^{2}, \\ 
C_{15}=-0.93+0.034P+0.014P^{2}, \\ 
C_{23}=21.11+3.69P-0.056P^{2}, \\ 
C_{25}=-0.70-0.10P-0.0086P^{2}, \\ 
C_{35}=-3.43+0.32P-6 \times 10^{-4}P^{2}, \\ 
C_{46}=-4.00-0.35P-0.0045P^{2},%
\end{array}%
\right.   \label{11} 
\end{eqnarray}
\begin{eqnarray}
&&\left\{ 
\begin{array}{c}
C_{11}=62.62+6.75P-0.10P^{2}, \\ 
C_{22}=57.74+7.95P-0.067P^{2}, \\ 
C_{33}=56.13+8.37P-0.12P^{2}, \\ 
C_{44}=17.94+1.90P-0.05P^{2}, \\ 
C_{55}=26.33+2.22P-0.026P^{2}, \\ 
C_{66}=17.08+1.98P-0.032P^{2},%
\end{array}%
\right. \left\{ 
\begin{array}{c}
C_{12}=18.19+2.60P-0.021P^{2}, \\ 
C_{13}=29.20+4.92P-0.054P^{2}, \\ 
C_{15}=-0.15+0.189P+0.014P^{2}, \\ 
C_{23}=19.41+3.15P-0.034P^{2}, \\ 
C_{25}=-1.64-0.34P-0.0036P^{2}, \\ 
C_{35}=-3.7-0.28P-1.4\times 10^{-3}P^{2}, \\ 
C_{46}=-3.7-0.28P-0.006P^{2},%
\end{array}%
\right.  
\end{eqnarray} 
\begin{eqnarray}
&&\left\{ 
\begin{array}{c}
C_{11}=62.43+7.15P-0.10P^{2}, \\ 
C_{22}=60.63+9.68P-0.15P^{2}, \\ 
C_{33}=62.016+7.78P-0.076P^{2}, \\ 
C_{44}=16.17+1.96P-0.052P^{2}, \\ 
C_{55}=27.36+3.62P-0.099P^{2}, \\ 
C_{66}=15.56+1.89P-0.036P^{2},%
\end{array}%
\right. \left\{ 
\begin{array}{c}
C_{12}=14.74+2.31P+0.002P^{2}, \\ 
C_{13}=28.91+4.39P-0.028P^{2}, \\ 
C_{15}=-0.73-0.38P+0.0079P^{2}, \\ 
C_{23}=13.48+2.903P-0.028P^{2}, \\ 
C_{25}=-2.87-0.091P-0.024P^{2}, \\ 
C_{35}=-1.59+1.19P-5.5\times 10^{-3}P^{2}, \\ 
C_{46}=-3.23-0.03P-0.012P^{2}.%
\end{array}%
\right. 
\end{eqnarray}

\vspace{10mm}
\subsubsection{Elastic constants for polycrystalline aggregates}
\vspace{6mm}
The three pairs of isotropic elastic parameters such as bulk modulus $B$ with
the shear modulus $G$ or the modulus of Young $E$ with Poisson's ratio $\delta $
or both Lam\'{e}'s constants $\lambda $ and $\mu $ can be used to fully
describe the mechanical behavior of a polycrystalline material.

The elastic constants $C_{ijs}$ in our paper have been estimated from \textsl{ab initio} PP-PW calculations for Ba$_{2}$P$_{7}$X monocrystalline. These
elastic constants $C_{ij}$ of the single-crystal are used to obtain the isotropic
elastic parameters. The mass modulus $B$, which measures the resistance of the
solid to the volume changes under the applied hydrostatic pressure. The
isotropic shear modulus $G$ is a measure of resistance to reversible
deformations caused by deformation. The shear strain can be determined
experimentally on a polycrystalline sample to characterize its mechanical
properties. Theoretically, $B$ and $G$ of the material's polycrystalline phase
can be obtained from the appropriate average of the independent elastic
constants $C_{ijs}$ of its monocrystalline phase. The modulus of elasticity
averaged by orientation $B$ and $G$ can be calculated using the Reuss-Voigt-Hill
approximations \cite{Voigt28,Hill52}. Here, the Voigt ($B_\text{V}$, $G_\text{V}$) and
Reuss ($B_\text{R}$,~$G_\text{R}$) approximations represent extreme values for $B$ and
$G$; and are expressed as follows \cite{Wu07}:
\begin{equation}
\left\{ 
\begin{array}{c}
B_\text{R}=\Omega \left[ 
\begin{array}{c}
a+\left( C_{11}+C_{22}-2C_{12}\right) +b\left( 2C_{12}-2C_{11}-C_{23}\right)
+c\left( C_{15}-2C_{25}\right) + \\ 
d\left( 2C_{12}+2C_{23}-C_{13}-2C_{22}\right) +2e\left( C_{25}-C_{15}\right)
+f%
\end{array}%
\right] ^{-1}, \\ 
B_\text{V}=\frac{1}{9}\left[ C_{11}+C_{22}+C_{33}+2\left(
C_{12}+C_{13}+C_{23}\right) \right],  \\ 
G_\text{V}=\frac{1}{15}\left[ C_{11}+C_{22}+C_{33}+3\left(
C_{44}+C_{55}+C_{66}\right) -\left( C_{12}+C_{13}+C_{23}\right) \right],  \\ 
G_\text{R}=15\left\{ \frac{4\left[ a\left( C_{11}+C_{22}+C_{12}\right) +b\left(
C_{11}-C_{12}-C_{23}\right) +c\left( C_{15}+C_{25}\right) +d\left(
C_{22}-C_{12}-C_{23}-C_{13}\right) +e\left( C_{15}-C_{25}\right) +f\right] }{%
\Omega +3\left[ \frac{g}{\Omega }+\left( \frac{C_{44}+C_{66}}{%
C_{44}C_{66}-C_{46}^{2}}\right) \right] }\right\} ^{-1}, \\ 
\text{with}: \\ 
a=C_{33}C_{55}-C_{35}^{2}\,, \\ 
b=C_{23}C_{55}-C_{25}C_{35}\,, \\ 
c=C_{13}C_{35}-C_{15}C_{33}\,, \\ 
d=C_{13}C_{55}-C_{15}C_{35}\,, \\ 
e=C_{13}C_{25}-C_{15}C_{23}\,, \\ 
f=C_{11}\big( C_{22}C_{33}-C_{25}^{2}\big) -C_{12}\left(
C_{12}C_{55}-C_{15}C_{25}\right) +C_{15}\left(
C_{12}C_{25}-C_{15}C_{22}\right) +C_{25}\left(
C_{23}C_{35}-C_{25}C_{33}\right), \\ 
g=C_{11}C_{22}C_{33}-C_{11}C_{23}^{2}-C_{22}C_{13}^{2}-C_{33}C_{12}^{2}+2C_{12}C_{13}C_{23}\,,
\\ 
\Omega =\left\{ 
\begin{array}{c}
2\left[ C_{15}C_{25}\left( C_{33}C_{12}-C_{13}C_{23}\right)
+C_{15}C_{35}\left( C_{22}C_{13}-C_{12}C_{23}\right) +C_{25}C_{35}\left(
C_{11}C_{23}-C_{12}C_{23}\right) \right] - \\ 
\left[ C_{15}^{2}\left( C_{22}C_{33}-C_{23}^{2}\right) +C_{25}^{2}\left(
C_{11}C_{33}-C_{13}^{2}\right) +C_{35}^{2}\left(
C_{11}C_{22}-C_{12}^{2}\right) \right] +gC_{55}%
\end{array}%
\right\} .
\end{array}%
\right.    \label{12}
\end{equation}

Hill recommends that the arithmetic mean of these two limits (Voigt, Reuss)
should be used in practice as an effective module for polycrystalline samples.
\begin{equation}
\left\{ 
\begin{array}{c}
B_\text{H}=\frac{B_\text{V}+B_\text{R}}{2}, \\ 
G_\text{H}=\frac{G_\text{V}+G_\text{R}}{2}.%
\end{array}%
\right.   \label{13}
\end{equation}
Where $B_\text{H}$ and $G_\text{H}$ are the shear modulus of the polycrystalline
according to Hill approximation. The Young's modulus $E$ and Poisson's ratio $%
\delta $ for an isotropic material can be computed from the Hill's values of
$B_\text{H}$ and $G_\text{H}$ using the following expressions,
\begin{equation}
\left\{ 
\begin{array}{c}
E=\frac{9B_\text{H}G_\text{H}}{3B_\text{H}+G_\text{H}}, \\ 
\delta =\frac{3B_\text{H}-2G_\text{H}}{6B_\text{H}+2G_\text{H}}.%
\end{array}%
\right.   \label{14}
\end{equation}

The calculated bulk modulus $B_\text{H}$, shear modulus $G_\text{H}$, Young's
modulus $E$ and Poisson's ratio $\delta $ are quoted in table~\ref{table_4}. The obtained
results allow us to make the following conclusions:

(i) From tables~\ref{table333} and \ref{table_4}, one can see that the value of the bulk modulus for
Ba$_{2}$P$_{7}$X deduced from the single-crystal elastic constants $C_{ijs}$ is
in good agreement with those calculated from the third order
polynomial $P(V)$, Birch-Murnaghan $P(V)$ EOS, Vinet $P(V)$ EOS, Murnaghan $P(V)$
EOS, Birch-Murnaghan $E(V)$ EOS and Murnaghan $E(V)$ EOS fits (figure~\ref{fig-smp3}). This
similarity may serve as an estimate of the reliability and accuracy of this
theoretical estimation of the elastic constants for the monoclinic Zintl
phase Ba$_{2}$P$_{7}$X. We can see that the bulk modulus of the considered
material is quite small (lower than 50 GPa), and therefore, this material
should be classified as a relatively soft material with high compressibility
(higher than 0.02) \cite{Haddadi10}.

(ii) The Young's modulus, which is defined to be the ratio of linear stress
to linear strain, may give information as to the stiffness of the material.
The Young's modulus of Ba$_{2}$P$_{7}$X was discovered to approximate 48 GPa
in the same order of $C_{11}$, $C_{22}$ and $C_{33}$ values, which indicates
the relatively high resistance of this compound to uniaxial deformation
(compression/traction); thus, these compounds show a rather low stiffness.
The highest Young's modulus belongs to Ba$_{2}$P$_{7}$Cl compound.
Therefore, this compound is harder than the other compounds.

(iii) The Poisson's ratio is the factor that measures the stability of a
crystal against shear \cite{Guechi14}, defined as the ratio of transverse
strain (normal to the applied stress) to the longitudinal strain (in the
direction of the applied stress), is generally connected with the volume
change in a solid during uniaxial deformation and provides more information
on the characteristics of the bonding forces than any of the other
elastic constants \cite{Haddadi10,Ravindran98,Bouhemadou13}. If $\delta $ is
equal to 0.5, no volume change occurs, while if it is lower than 0.5, a large
volume change is expected for any elastic deformation \cite{Appalakondaiah12}%
, and it has been proved that $\delta $ =0.25 is the lower limit for central
force and $\delta $ = 0.5 is the upper limit. In our case, the value of $%
\delta $ is approximately 0.30 in Ba$_{2}$P$_{7}$X, suggesting that a
considerable volume change can be associated with elastic deformation and
that the interatomic forces in this compound are central.

(vi) The bulk and shear moduli provide information regarding the
brittle-ductile nature of a material. Pugh \cite{Pugh54} has proposed a
simple empirical relationship between the bulk modulus $B_\text{H}$ and shear
modulus $G_\text{H}$. According to these criteria, the calculated value of $B_\text{H}/G_\text{H}$ is higher
than 1.75, which may be associated to the ductility; whereas a value lower than 1.75 is
associated to brittleness. Based on Pugh's criteria, the three considered compounds Ba%
$_{2}$P$_{7}$X are ductile materials.

(v) The Debye temperature $\theta _\text{D}$ is used to distinguish between high
and low temperatures for a solid in the Debye model. The Debye temperature $%
\theta_\text{D}$ is correlated with many physical properties, such as thermal
expansion, melting point and Gr\"{u}neisen parameter. The flow temperature $%
\theta_\text{D}$ can be estimated numerically from the mean speed of the sound
wave $V_\text{m}$ as follows \cite{Anderson63}:
\begin{equation}
\theta _\text{D}=\frac{h}{k_\text{B}}V_\text{m}\left[ \frac{3n}{4\piup }\frac{N_\text{A}\rho }{M}%
\right] ^{1/3}.  \label{15}
\end{equation}
where, $h$ is Planck constant, $k_\text{B}$ is Boltzmann constant, $N_\text{A}$ is
Avogadro number, $\rho $ is the mass density, M is the molecular weight and $%
n $ is the number of atoms in the molecule. In polycrystalline materials, the
average wave velocity $V_\text{m}$ can be evaluated as follows \cite{Anderson63}:
\begin{equation}
V_\text{m}=\left[ \frac{1}{3}\left( 2V_\text{T}^{-3}+V_\text{L}^{-3}\right) \right] ^{-1/3}.
\label{16}
\end{equation}
Here, $V_\text{L}$ and $V_\text{T}$ are the average longitudinal and transverse
elastic wave velocities, which are defined by Navier's equations  \cite%
{Schreiber74}:
\begin{equation}
V_\text{L}=\left( \frac{3B+4G}{3\rho }\right) ^{1/2},  \label{17}
\end{equation}
\begin{equation}
V_\text{T}=\left( \frac{G}{\rho }\right) ^{1/2}.  \label{18}
\end{equation}

The results obtained for $\theta _\text{D}$ and average sound velocities are
listed in table~\ref{table_4}. The sound velocity in the Ba$_{2}$P$_{7}$Cl compound
is higher in the two compounds Ba$_{2}$P$_{7}$Br and Ba$_{2}$P$_{7}$I, like
Debye temperatures ($\theta _\text{D}$) values. Therefore, it can be said that
the sound conductivity of Ba$_{2}$P$_{7}$Cl is better than that of other
compounds.

\begin{table}[!t]%
\caption{The calculated bulk modulus ($B_\text H$, in GPa); shear modulus ($G_\text H$, in GPa); Young's modulus ($E$, in GPa); 
Poisson's ratio ($\delta$); mass density $\rho$ (g/cm$^3$); longitudinal, transverse and average sound velocities 
($V_\text L$, $V_\text T$ and $V_\text m$, in m/s) and Debye temperature ($\theta_\text D$, in K). For the monoclinic Zintl phase Ba$_2$P$_7$X, it is obtained using the single-crystal elastic constants $C_{ijs}$. 
The subscript V, R or H indicates that the modulus was obtained using Voigt theory or Reuss theory and Hill theory,
 respectively.}
 \vspace{2ex}
 \renewcommand{\arraystretch}{1.2}
 \begin{center}
\begin{tabular}[b]{|c|c|c|c|c|c|c|c|}
\hline
System & $\rho $ & $B_\text{V}$ & $B_\text{R}$ & $B_\text{H}$ & $G_\text{V}$ & $G_\text{R}$ & $G_\text{H}$
\\ \hline
Ba$_{2}$P$_{7}$Cl & 3.58 & 34.22 & 33.80 & 34.01 & 19.44 & 18.26 & 18.85 \\ 
\hline
Ba$_{2}$P$_{7}$Br & 3.83 & 34.45 & 35.05 & 34.75 & 19.50 & 17.95 & 18.55 \\ 
\hline
Ba$_{2}$P$_{7}$I & 3.75 & 33.18 & 37.21 & 34.20 & 20.36 & 16.70 & 18.53 \\ 
\hline
System & $B_\text{H}/G_\text{H}$ & $E$ & $\delta $ & $V_\text{L}$ & $V_\text{T}$ & $V_\text{m}$ & $%
\theta _\text{D}$ \\ \hline
Ba$_{2}$P$_{7}$Cl & 1.80 & 47.74 & 0.266 & 4064.91 & 2294.92 & 2552.66 & 262
\\ \hline
Ba$_{2}$P$_{7}$Br & 1.87 & 47.25 & 0.273 & 3941.31 & 2200.95 & 2450.31 & 250
\\ \hline
Ba$_{2}$P$_{7}$I & 1.89 & 47.30 & 0.276 & 3997.06 & 2149.86 & 2394.20 & 242
\\ \hline
\end{tabular}%
 \end{center}
\label{table_4}
\end{table}%

\begin{figure}[htb]
\includegraphics[width=0.49\textwidth]{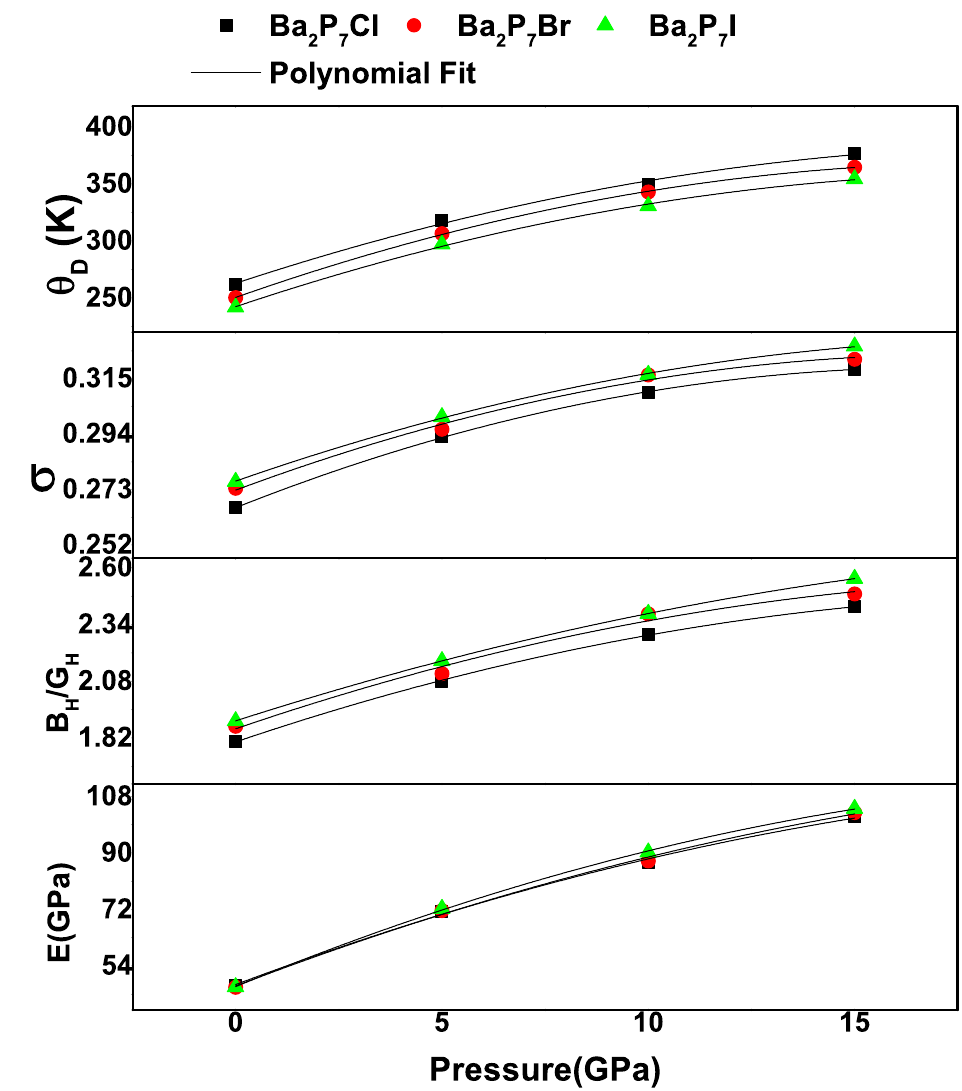}
\includegraphics[width=0.49\textwidth]{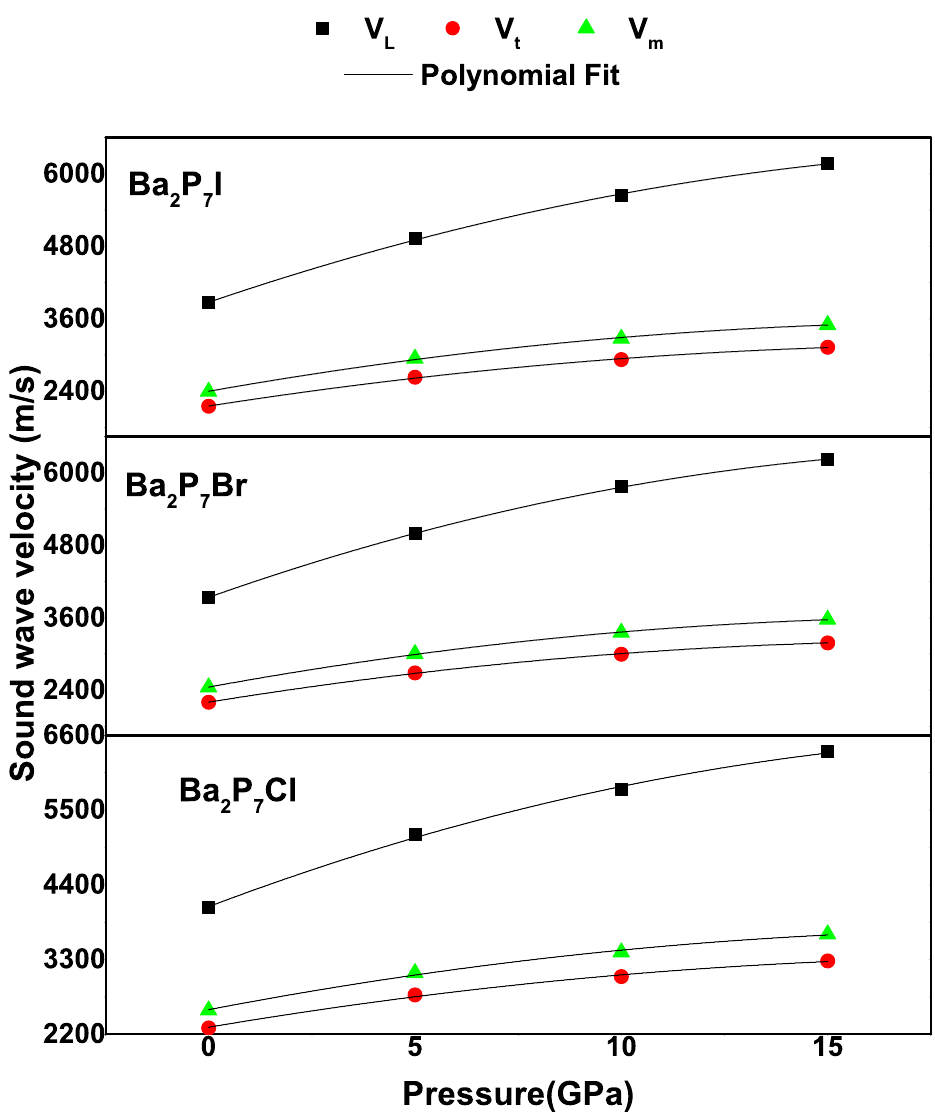}
\caption{(Colour online) The calculated pressure dependence of Pugh's ratio $B_\text H/G_\text H$, Young's modulus~$E$, Poisson's ratio $\delta$, Debye temperature $\theta_\text D$, and the isotropic sound velocity (longitudinal $V_\text L$, transverse $V_\text T$ and average $V_\text m$) 
 for the monoclinic Zintl phase Ba$_2$P$_7$X. The symbols indicate the calculated results.
 The lines represent the results of fitting these theoretical results to a second-order polynomial.} \label{fig-smp5}
\end{figure}

From this table, it can be seen that Ba$_{2}$P$_{7}$X is characterized by a
high Debye temperature equal to 260~K, the Debye temperature, the behavior
of the sound velocity, the Young's modulus $E$, the Poisson's ratios $\delta $
and the Pugh $B_\text{H}/G_\text{H}$ ratio under the effect of pressure are shown
in figure~\ref{fig-smp5}, this figure shows that all these physical parameters increase
with an increase of the pressure and are well adjusted by a second order
polynomial equation for the three compounds Ba$_{2}$P$_{7}$CL, Ba$_{2}$P$_{7}$%
Br, Ba$_{2}$P$_{7}$I, respectively:

\begin{eqnarray}
&&\left\{ 
\begin{array}{lll}
\frac{B_\text{H}}{G_\text{H}}=1.8+0.064P-1.52\times 10^{-3}P^{2}, \\ 
E=48.08+4.94P-0.093P^{2}, \\ 
\delta =0.266+6.2\times10^{-3}P-1.8\times10^{-4}P^{2}, \\ 
V_\text{L}=4078.73+228.88P-5.21P^{2}, \\ 
V_\text{T}=2303.80+102.27P-2.52P^{2}, \\ 
V_\text{m}=2562.5+116.06P-2.85P^{2}, \\ 
\theta _\text{D}=262.93+11.91P-0.293P^{2}, %
\end{array}%
\right. {\small }  \label{19} 
\end{eqnarray}
\newpage
\begin{eqnarray}
&&\left\{ 
\begin{array}{lll}
\frac{B_\text{H}}{G_\text{H}}=1.86+0.0646P-1.52\times 10^{-3}P^{2}, \\ 
E=47.65+5.07P-0.09P^{2}, \\ 
\delta =0.272+5.8\times10^{-3}P-1.64\times 10^{-4}P^{2}, \\ 
V_\text{L}=3938.99+241.2P-5.91P^{2}, \\ 
V_\text{T}=2203.42+110.12P-2.98P^{2}, \\ 
V_\text{m}=2452.8+124.65P-3.35P^{2}, \\ 
\theta _\text{D}=250.55+12.73P-0.34P^{2},%
\end{array}%
\right. {\small } \\
&&\left\{ 
\begin{array}{lll}
\frac{B_\text{H}}{G_\text{H}}=1.89+0.060P-1.15\times10^{-3}P^{2}, \\ 
E=47.44+5.43P-0.11P^{2}, \\ 
\delta =0.276+5.4\times10^{-3}P-1.37\times10^{-4}P^{2}, \\ 
V_\text{L}=3873.63+232.73P-5.34P^{2}, \\ 
V_\text{T}=2155.03+106.2P-2.78P^{2}, \\ 
V_\text{m}=2399.7+120.67P-3.17P^{2}, \\ 
\theta _\text{D}=242.45+12.13P-0.315P^{2}.%
\end{array}%
\right. 
\end{eqnarray}

\subsubsection{Elastic anisotropy}

The anisotropy of the physical properties in the crystals and a correct
description of the anisotropic behavior, and the elastic anisotropy is
another interesting physical parameter with respect to the elastic
properties of the solids. It reflects the anisotropy in the bond between the
atoms in different crystallographic directions. Anisotropic characters of
binding and structural stability are usually defined by the elastic
constants $C_{ijs}$. These constants have been often related to the shear modulus
$G$ and Young's modulus $E$. We have previously reported that the three Ba$_{2}$P%
$_{7}$X materials are anisotropic in terms of compressibility (figure~\ref{fig-smp2}). It
is important to evaluate the elastic anisotropy of a solid to understand the
micro cracks that are easily induced in materials due to a significant
anisotropy of the coefficient of thermal expansion as well as the elastic
anisotropy \cite{Ravindran98} and its influence on nanoscale precursor
textures of alloys \cite{Lloveras08}. Different approaches were developed to
describe the materials' elastic anisotropy. Four different criteria were employed to quantify the anisotropy of the
elastic properties of the Ba$_{2}$P$_{7}$Cl, Ba$_{2}$P$_{7}$Br and Ba$_{2}$P$%
_{7}$I compounds.

(i) A method of measuring the elastic anisotropy which consists in
considering the percentage of anisotropy in the compression and shear modulus
was proposed by Chung and Buessem \cite{Chung68}:

\begin{equation}
\left\{ 
\begin{array}{c}
A_{B}=\frac{B_\text{V}-B_\text{R}}{B_\text{V}+B_\text{R}}\times 100\,, \\ 
A_{G}=\frac{G_\text{V}-G_\text{R}}{G_\text{V}+G_\text{R}}\times 100\,,
\end{array}%
\right.   \label{20}
\end{equation}
\begin{table}[!b]%
	\caption{The calculated percentage of elastic anisotropy for bulk modulus and
		shear modulus ($A_\text B$ and $A_\text G$) and universal anisotropy index ($A^\text U$) for
		the Ba$_2$P$_7$Cl, Ba$_2$P$_7$Br and Ba$_2$P$_7$I compounds.}%
	\vspace{2ex}
	\renewcommand{\arraystretch}{1.2}
	\begin{center}
		\begin{tabular}[b]{|c|c|c|c|}
			\hline
			System & $A_\text{B}$ \% & $A_\text{G}$ \% & $A^\text{U}$ \% \\ \hline
			Ba$_{2}$P$_{7}$Cl & 0.61 & 3.12 & 0.33 \\ \hline
			Ba$_{2}$P$_{7}$Cl & 0.85 & 10.86 & 0.52 \\ \hline
			Ba$_{2}$P$_{7}$Cl & 5.73 & 9.85 & 0.98 \\ \hline
		\end{tabular}%
	\end{center}
	\label{table5}
	\vspace{-2mm}
\end{table}%
where $B$ and $G$ are the bulk and shear moduli, respectively, and the
subscripts V and R represent the Voigt and Reuss bounds, a value of
zero (0 \%) represents elastic isotropy and a value of (100~\%) represents the
largest possible elastic anisotropy. The results shown in table~\ref{table5} for $A_{B}$
and $A_{G}$ suggest that Ba$_{2}$P$_{7}$Cl, Ba$_{2}$P$_{7}$Br and Ba$_{2}$P$%
_{7}$I compounds are anisotropic.

(ii) A universal anisotropy index $A^\text{U}$ was proposed by Ranganathan and
Ostoja-Starzewski \cite{Ranganathan08} to quantify the elastic anisotropy of
three crystals accounting for bulk and shear modulus contributions. The
index $A^\text{U}$ is delimited as follows:
\begin{equation}
A^\text{U}=5\frac{G_\text{V}}{G_\text{R}}+\frac{B_\text{V}}{B_\text{R}}-6.  \label{21}
\end{equation}

For isotropic crystals, the universal index is equal to zero ($A^\text{U}=0$);
the deviation of $A^\text{U}$ from zero defines the extent of the anisotropy of
a crystal. The results listed in table~\ref{table5} for $A^\text{U}$ indicate that Ba$_{2}$%
P$_{7}$Cl, Ba$_{2}$P$_{7}$Br and Ba$_{2}$P$_{7}$I have a certain degree of
elastic anisotropy.

\section{Conclusions}

In this paper, a prediction of some physical properties of the monoclinic
Zintl phase Ba$_{2}$P$_{7}$X (X=Cl, Br, I) was obtained by using the PP-PW
method based on DFT with the GGA PBEsol approach. At first, an accurate
geometrical optimization was performed on the crystal structure, and
then the structural and elastic properties of the three materials Ba$_{2}$P$%
_{7}$Cl, Ba$_{2}$P$_{7}$Br and Ba$_{2}$P$_{7}$I were calculated in
detail respectively. The results show that:

\textbullet \qquad The theoretically predicted lattice parameters for Ba$%
_{2} $P$_{7}$Cl, Ba$_{2}$P$_{7}$Br and Ba$_{2}$P$_{7}$I are in good
agreement with the existing experimental measurements. The calculated
zero-pressure single-crystal elastic constants Cijs of Ba$_{2}$P$_{7}$X
(X=Cl, Br, I) satisfy the dynamical stability criteria.

\textbullet \qquad The pressure dependence of the elastic constants reveals
that Ba$_{2}$P$_{7}$X (X=Cl, Br, I) remains mechanically stable under
hydrostatic pressure effect as well.

\textbullet \qquad This paper calculates and estimates the elastic
constants, and other related quantities consisting in Young's modulus, shear
modulus, Poisson's ratio, anisotropy factor, sound velocities, and Debye
temperature.

\textbullet \qquad The material has a relatively small bulk modulus and a
brittle character. The bulk modulus derived from the single-crystal elastic
constants $C_{ijs}$ is observed to be in excellent agreement with the one
estimated from the EOS-fitting. This result shows the reliability of our
calculations.

\textbullet \qquad The investigated properties demonstrate that the three
compounds are relatively soft materials.

\textbullet \qquad The elastic constants of three single-crystal and
polycrystalline phases of Ba$_{2}$P$_{7}$X were estimated. The Ba$_{2}$P$%
_{7} $X compounds exhibit a noticeable elastic anisotropy. And finally, by
using the empirical rule of Pugh, the $B/G$ ratio, we have demonstrated that
the studied compound should be classified as a relatively ductile
material.

\section*{Acknowledgements}

The authors would like to thank Dr. Zitouni H., Dr. Ahmed Ammar M., Mr. Media M. and Mr.~Houatis~D. for their, help, support, and constant assistance and for their advice throughout this project.

\newpage

\ukrainianpart

\title{Дослідження структурних та пружних властивостей моноклінних Ba$_2$P$_7$X (X $=$ Cl, Br, I) солей Цінтля}
\author{M. Раджай\refaddr{label1}\refaddr{label2}, Д. Мауш\refaddr{label1}, Н. Гуеші\refaddr{label3}, С. Шеддаді\refaddr{label4}, З. Кешіді \refaddr{label5}}
\addresses{
	\addr{label1} Лабораторія для отримання нових матеріалів і їхня характеризація, університет Ферхат Аббас Сетіф 1, Алжир
	\addr{label2} Лабораторія фізики експериментальних методик та їх застосування   (LPTEAM), університет м. Медеа, Алжир
	\addr{label3} Лабораторія досліджень поверхонь та інтерфейсів твердих матеріалів, 
	університет Ферхат Аббас Сетіф 1, Алжир
	\addr{label4} Лабораторія радіаційної фізики, фізичний відділ, факультет природничих наук, університет Ваджі Мохтар, Аннаба, Алжир
	\addr{label5} Лабораторія електротехніки та автоматики LREA, університет м. Медеа, Алжир
}

\makeukrtitle

\begin{abstract}
	 Досліджувалися структурні і пружні властивості  Ba$_{2}$P$_{7}$X (X$=$Cl, Br, I)  сполук Цінтля, використовуючи метод псевдопотнціальної плоскої хвилі (PP-PW), що базується на теорії функціоналу густини (DFT)  в межах узагальненого градієнтного наближення (GGA-PBESOL). Розраховані сталі гратки і внутрішні параметри добре узгоджуються з експериментальними результатами, відомими з літератури. В цій статті ми представляємо дослідження відносних змін структурних параметрів і пружних констант як функцій гідростатичного тиску. Ізотропні  модулі пружності та пов'язані з ними властивості для монокристала і полікристалічної фази, включаючи, зокрема, об'ємні модулі, зсувні модулі, модулі Юнга, коефіцієнт Пуассона, пружні анізотропні індекси, індикатор   Пуга  поведінки крихкість/пластичність, швидкості пружної хвилі і температура Дебая були оцінені з $C_{ij}$, використовуючи наближення Войгта, Реусса і Хілла. Два різні методи були використані для вивчення пружної анізотропії цих сполук.
	\keywords сполука Цінтля, P$_{7}^{-3}$ кластери, модуль пружності,
	\textsl{ab initio} обчислення
	
\end{abstract}


\begin{thebibliography}{99}
\bibitem{Dolyniuk13} Dolyniuk J.-A., Kovnir K., Crystals, 2013, \textbf{3}, 431--442, \doi{10.3390/cryst3030431}.

\bibitem{Eschen02} Eschen M., Jeitschko W., J. Solid State Chem.,
2002, \textbf{165}, 238--246, \doi{10.1006/jssc.2001.9497}.

\bibitem{Kraus11} Kraus F., Korber N., Chem. Eur. J., 2005, \textbf{11}, 5945--5959, \doi{10.1002/chem.200500414}.

\bibitem{Dell98} Dell S., Vogelaar N.J., Ho D.M., Pascal
R.A., J. Am. Chem. Soc., 1998, \textbf{120}, 6421--6422, \doi{10.1021/ja981009s}.

\bibitem{Miller11} Miller G.J., Schmidt M.W., Wang F., You T.-S., In: Zintl Phases. Structure and Bonding, Vol.~139, F\"{a}ssler T. (Ed.), Springer, Berlin, Heidelberg, 2011, 1--55, \doi{10.1007/430_2010_24}.

\bibitem{Manriquez86} Manriquez V., H\"{o}nle W., von Schnering H.G., Z. Anorg. Allg. Chem., 1986, \textbf{539}, 95--109, \\ \doi{10.1002/zaac.19865390810}.

\bibitem{Dahlmann73} Dahlmann W., von Schnering H.G., Naturwissenschaften,
1973, \textbf{60}, 518, \doi{10.1007/BF00603256}.

\bibitem{Sin'ko08} Sin'ko G.V., Phys. Rev. B, 2008, \textbf{77}, 104118, \doi{10.1103/PhysRevB.77.104118}.

\bibitem{Zhijiao11} Zhijiao Z., Feng W., Zhou Z., Jianjun W., Xinyou A., Guo
L., Weiyi R., Physica B, 2011, \textbf{406}, 737, \\ \doi{10.1016/j.physb.2010.11.040}.

\bibitem{Clark05} Clark S.J., Segall M.D., Pickard C.J., Hasnip P.J.,
Probert M.J., Refson K., Payne M.C., Z. Kristallogr., 2005, \textbf{220}, 567, \doi{10.1524/zkri.220.5.567.65075}.

\bibitem{Perdew08} Perdew J.P., Ruzsinszky A., Csonka G.I., Vydrov O.A., Scuseria G.E., Constantin L.A., Zhou X., Burke K., Phys. Rev. Lett., 2008,
\textbf{100}, 136406, \doi{10.1103/PhysRevLett.100.136406}.

\bibitem{Vanderbilt92} Vanderbilt D., Phys. Rev. B, 1990, \textbf{41}, 7892(R), \doi{10.1103/PhysRevB.41.7892}.

\bibitem{Monkhorst76} Monkhorst H.J., Pack J.D., Phys. Rev. B,  1976, \textbf{13},
5188, \doi{10.1103/PhysRevB.13.5188}. 

\bibitem{Voigt28} Voigt W., Lehrbuch der Kristallphysik (Textbook of Crystal
Physics), Teubner, Leipzig, 1928.

\bibitem{Hill52} Hill R.,
Proc. Phys. Soc. London, Sect. A, 1952, \textbf{65}, 349--354, \doi{10.1088/0370-1298/65/5/307}.

\bibitem{Schnering81} Schnering H.G.V., Menge G.,
 Z. Anorg. Allg. Chem., 1981, \textbf{481},
33--40, \doi{10.1002/zaac.19814811005}. 

\bibitem{Wu10} Wu M.-M., Wen L., Tang B.-Y., Peng L.-M., Ding W.-J., J.
Alloys Compd., 2010, \textbf{506}, 412, \\ \doi{10.1016/j.jallcom.2010.07.018}.

\bibitem{Ambrosch-Draxl06} Ambrosch-Draxl C., Sofo J.O., Comput. Phys. Commun., 2006, \textbf{175}, 1--14, \doi{10.1016/j.cpc.2006.03.005}.

\bibitem{Hebbache04} Hebbache M., Zemzemi M., Phys. Rev. B,
2004, \textbf{70}, 224107, \doi{10.1103/PhysRevB.70.224107}.

\bibitem{Birch78} Birch F., J.
Geophys. Res., 1978, \textbf{83}, 1257--1268, \doi{10.1029/JB083iB03p01257}.

\bibitem{Fu83} Fu C.-L., Ho K.-M., Phys. Rev. B, 1983, \textbf{28}, 5480, \doi{10.1103/PhysRevB.28.5480}.

\bibitem{Murnaghan44} Murnaghan F.D., Proc. Natl. Acad. Sci. U.S.A., 1944, \textbf{30}, 244--247, \doi{10.1073/pnas.30.9.244}.

\bibitem{Birch47} Birch F., Phys.
Rev., 1947, \textbf{71}, 809, \doi{10.1103/PhysRev.71.809}.

\bibitem{Westbrook00} Westbrook J.H., Fleischer R.L., John Wiley
\& Sons Ltd, Baffins Lane, Chichester, West Sussex PO 19 IUD, England,  2000.

\bibitem{Wu07} Wu Z.-J., Zhao E.-J., Xiang H.-P., Hao X.-F., Liu X.-J., Meng J.,
Phys. Rev. B, 2007, \textbf{76}, 054115, \\ \doi{10.1103/PhysRevB.76.054115}.

\bibitem{Sin'ko02} Sin'ko G.V., Smirnov N.A., J. Phys.: Condens.
Matter, 2002, \textbf{14}, 6989, \doi{10.1088/0953-8984/14/29/301}.

\bibitem{Haddadi10} Haddadi K., Bouhemadou A., Louail L., Solid State Commun., 2010, \textbf{150}, 932--937, \\ \doi{10.1016/j.ssc.2010.02.024}.

\bibitem{Guechi14} Guechi N., Bouhemadou A., Khenata R., Bin-Omran S.,
Chegaar M., Al-Douri Y., Bourzami A., Solid State Sci., 2014, \textbf{29}, 12--23, \doi{10.1016/j.solidstatesciences.2014.01.001}.

\bibitem{Ravindran98} Ravindran P., Fast L., Korzhavyi P.A., Johansson
B., J. Appl. Phys., 1998, \textbf{84}, 4891, \doi{10.1063/1.368733}.

\bibitem{Bouhemadou13} Bouhemadou A., U\u{g}ur G., U\u{g}ur \c{S}., Al-Essa
S., Ghebouli M.A., Khenata R., Bin-Omran S., Al-Dour Y., Comput. Mater. Sci., 2013, \textbf{70}, 107--113, \doi{10.1016/j.commatsci.2013.01.004}.

\bibitem{Appalakondaiah12} Appalakondaiah S., Vaitheeswarm G., Leb\`{e}gue
S., Christensen N.E., Svane A., Phys. Rev. B, 2012, \textbf{86}, 035105, \\ \doi{10.1103/PhysRevB.86.035105}.

\bibitem{Pugh54} Pugh S.F., Philos. Mag., 1954, \textbf{45}, 823--843, \doi{10.1080/14786440808520496}.

\bibitem{Anderson63} Anderson O.L., J. Phys. Chem. Solids, 1963, \textbf{24},
909--917, \doi{10.1016/0022-3697(63)90067-2}.

\bibitem{Schreiber74} Schreiber E., Anderson O.L., Soga N., Elastic
Constants and Their Measurements, McGraw-Hill Companies, New York, 1974.

\bibitem{Lloveras08} Lloveras P., Cast\'{a}n T., Porta M., Planes A., Saxena
A., Phys. Rev.
Lett., 2008, \textbf{100}, 165707, \\ \doi{10.1103/PhysRevLett.100.165707}.

\bibitem{Chung68} Chung D.H., Buessem W.R., In: Anisotropy in Single-Crystal Refractory Compounds, Vol.~2, Vahldiek F.W., Mersol~S.A.
(Eds.), Plenum, New York,
1968, 217--246, \doi{10.1007/978-1-4899-5307-0}.

\bibitem{Ranganathan08} Ranganathan S.I., Ostoja-Starzewski M., Phys. Rev. Lett., 2008, \textbf{101}, 055504, \\
\doi{10.1103/PhysRevLett.101.055504}.
\end{thebibliography}
\end{document}